\documentclass{aa}  
\usepackage{natbib}
\usepackage[colorlinks=true,citecolor=blue]{hyperref}
\usepackage{graphicx}
\usepackage{txfonts}
\begin{document}

   \title{The GTC exoplanet transit spectroscopy survey}

   \subtitle{XI. Possible detection of Rayleigh scattering in the atmosphere of the Saturn-mass planet WASP-69b}

  \author{F. Murgas
         \inst{1,2}
         \and
         G. Chen\inst{3}
         \and
         L. Nortmann \inst{1,2}
         \and
         E. Palle \inst{1,2}
         \and
         G. Nowak \inst{1,2}
         }
   \institute{Instituto de Astrof\'{i}sica de Canarias (IAC), 38205 La Laguna, Tenerife, Spain\\
              \email{fmurgas@iac.es}
         \and
             Departamento de Astrof\'{i}sica, Universidad de La Laguna (ULL), 38206 La Laguna, Tenerife, Spain
             \and
             Key Laboratory of Planetary Sciences, Purple Mountain Observatory, Chinese Academy of Sciences, Nanjing 210023, China
             }
   \date{Received 14 April 2020/ Accepted 3 July 2020}

 
  \abstract
   {One of the main atmospheric features in exoplanet atmospheres, detectable both from ground- and space-based facilities, is Rayleigh scattering. In hydrogen-dominated planetary atmospheres, Rayleigh scattering causes the measured planetary radius to increase toward blue wavelengths in the optical range.}
   {We aim to detect and improve our understanding of several features in the optical range observable in planetary atmospheres. We focus on studying transiting exoplanets that present a wide range of orbital periods, masses, radii, and irradiation from their host star.}
   {We obtained a spectrophotometric time series of one transit of the Saturn-mass planet WASP-69b using the OSIRIS instrument at the Gran Telescopio Canarias. From the data we constructed 19 spectroscopic transit light curves representing 20 nm wide wavelength bins spanning from 515 nm to 905 nm. We derived the transit depth for each curve individually by fitting an analytical model together with a Gaussian process to account for systematic noise in the light curves.}
   {We find that the transit depth increases toward bluer wavelengths, indicative of a larger effective planet radius. Our results are consistent with space-based measurements obtained in the near infrared using the Hubble Space Telescope, which show a compatible slope of the transmission spectrum. We discuss the origin of the detected slope and argue between two possible scenarios: a Rayleigh scattering detection originating in the planet's atmosphere or a stellar activity induced signal from the host star.
   }
   {}


   \maketitle
%

\section{Introduction}

An active field of exoplanet science is the characterization of atmospheres of transiting planets. Due to their favorable orbital configuration and chance alignment with our line of sight, transiting planets allow us to study their atmospheres to search for atmospheric features and probe their compositions. Transmission spectroscopy has become a successful tool for studying the upper layers of planetary atmospheres near the terminator region during transit (e.g., \citealp{2002ApJ...568..377C}, \citealp{2008ApJ...673L..87R}, \citealp{2011MNRAS.412.2376W}, \citealp{2012MNRAS.426.1663S}, \citealp{2017A&A...600L..11C}). In the past decade, such studies were conducted for a number of planets, many of which were revealed to possess cloudy and hazy atmospheres (e.g., \citealp{2011MNRAS.416.1443S}, \citealp{2014Natur.505...69K}). These cloudy-hazy planets reveal themselves in the form of either flat spectra, which are most likely explained by high cloud decks, or by sloped transmission spectra, with larger absorption toward shorter wavelengths (e.g., \citealp{2008A&A...481L..83L}, \citealp{2013ApJ...778..184J}). Coupled with muted molecular features, these sloped spectra are best explained by photochemical hazes in the atmosphere (e.g., \citealp{2016Natur.529...59S}) or by atmospheric Rayleigh scattering. While high-resolution transmission studies probe individual lines (e.g., \citealp{2017A&A...602A..36W}, \citealp{2018A&A...616A.151C}, \citealp{2019A&A...623A.166S}, \citealp{2019A&A...628A...9C}, \citealp{2020A&A...635A.171C}), studies at lower resolution can give us information about the continuum of the exoplanet spectrum (e.g., \citealp{2017ApJ...834..151R}, \citealp{2018AJ....156..122M}, \citealp{2019AJ....157...68B}, \citealp{2019A&A...631A.169T}, \citealp{2020AJ....159....7M}, \citealp{2020AJ....159...13W}). Moreover, ground- and space-based low-resolution transmission spectroscopy allows us to probe for potential broad absorption line wings and slopes caused by hazes in the atmosphere of Jovian-like transiting planets. The presence of broad absorption lines, slopes at optical wavelengths, and/or water absorption bands in the near infrared can help us break the degeneracy between the atmospheric pressure and abundances of elements using spectral retrieval analysis techniques (e.g., \citealp{2012ApJ...753..100B}, \citealp{2017MNRAS.470.2972H}). Hence, having a library of ground- or spaced-based optical and near-infrared observations can help establish abundances of transiting planets using current facilities like Hubble Space Telescope Wide Field Camera 3 (HST/WFC3), or complement future observations made with James Webb Space Telescope (JWST).

Here we report the analysis of one planetary transit of WASP-69b obtained with the OSIRIS instrument (Optical System for Imaging and low-Intermediate-Resolution Integrated Spectroscopy; \citealp{2012SPIE.8446E..4TS}) mounted on the 10.4 m Gran Telescopio Canarias (GTC). WASP-69b is a Saturn-mass (0.26 $M_{\mathrm{Jup}}$) highly inflated (1.06 $R_{\mathrm{Jup}}$) planet orbiting a K-type star with a period of approximately 3.9 days \citep{2014MNRAS.445.1114A}. This planet is of special interest due to its low mean density and due to the fact that it has been shown, through studies of its near-infrared helium signature, to lose its atmosphere in the form of a tail  \citep{2018Sci...362.1388N}. Using WFC3 on the Hubble Space Telescope (HST), the planet was found to have a muted water feature at 1.4 $\mu$m with an overlying slope of a larger planet radius toward bluer wavelengths \citep{2018AJ....155..156T}. Furthermore, absorption in Na D was detected during transit using high-resolution observations obtained with the HARPS spectrograph \citep{2017A&A...608A.135C}, although an independent study using ESPaDOnS/CFHT could not confirm the sodium detection \citep{2019AJ....157...58D}.

This paper is organized as follows. In Sect. \ref{sec:data} we present the observations and data reduction strategy. In Sect. \ref{sec:analysis} we describe our light curve fitting procedure. In Sects. \ref{sec:TransOptical} and \ref{sec:TransFull} we present our optical WASP-69b transmission spectrum and joint analysis of our results, in addition to a published near-infrared transmission spectrum of the planet. In Sect. \ref{sec:conclusions} we present the conclusions of this work.

\begin{table*}
 \caption[]{Coordinates of the planet-host star WASP-69 and the reference star used in the observing run. }
\begin{center}
\setlength{\tabcolsep}{1.5mm}
\begin{tabular}{c c c c c c}
\hline\hline
Star&2MASS  ID & RA $^*$&Dec $^*$& $V_{\rm mag}$ & $(B-V)_{\rm mag}$ \\ \hline
\hspace{20mm}\newline
WASP-69    & J21000618-0505398 & $21$h $00$m $05.20$s&$-05\degr$ $05\arcmin$ $40.037\arcsec$ & 9.87\tablefootmark{b} & 1.06\tablefootmark{b}\\ 
Reference  & J20595897-0507535 & $20$h $59$m $59.00$s&$-05\degr$ $07\arcmin$ $53.766\arcsec$ & 10.60\tablefootmark{b} & 1.03\tablefootmark{b} \\
 \hline
\end{tabular}
\tablefoot{
\tablefoottext{a}{Reference for coordinates: \cite{2018yCat.1345....0G}}
\tablefoottext{b}{\cite{2000A&A...355L..27H}}
}
\label{tab:coord}  
\end{center}
\end{table*}

\section{Data observations and reduction}
\label{sec:data}
\subsection{Observing setup}
\begin{figure}
   \centering
   \includegraphics[width=\hsize]{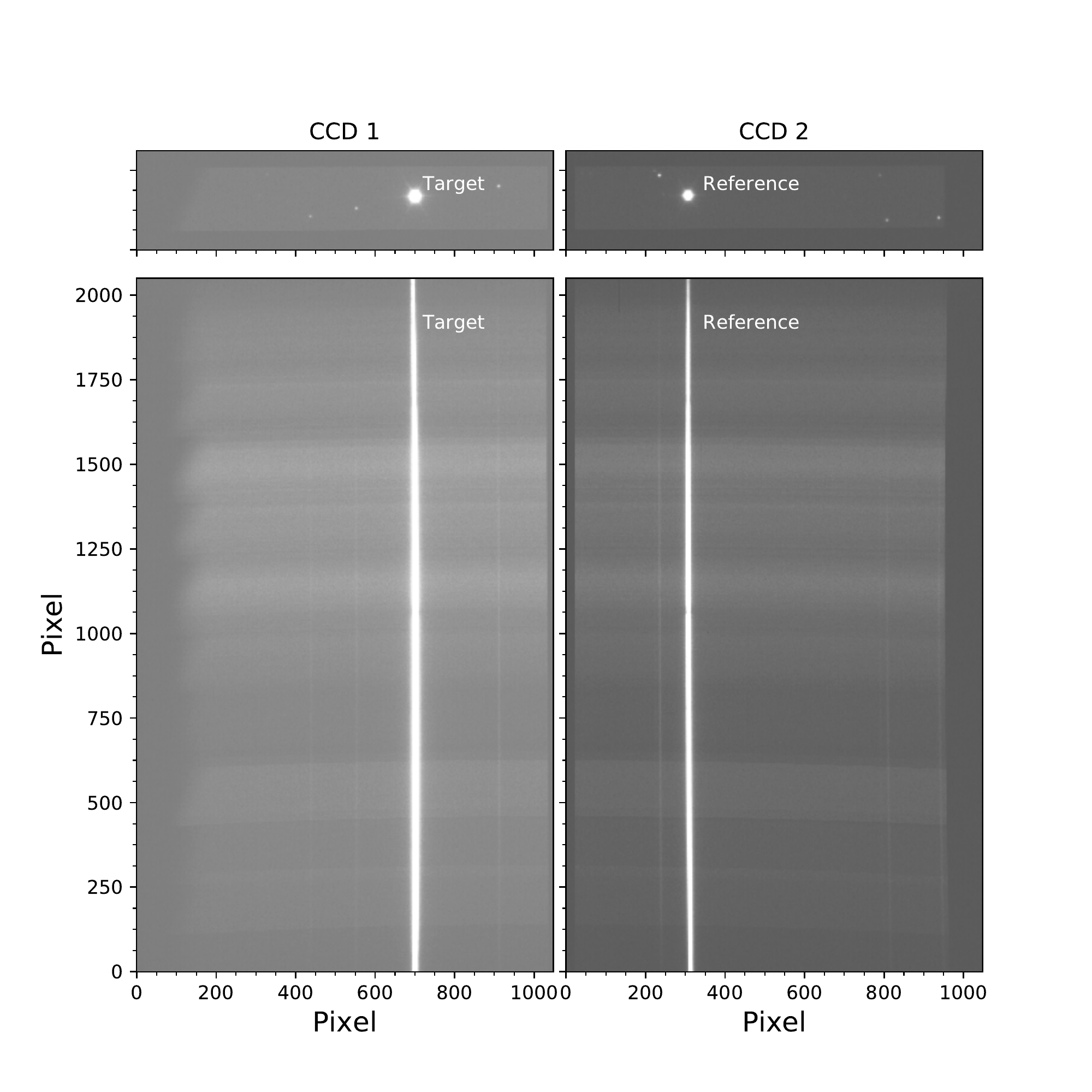}
   \caption{\textit{Top panel}: Through slit image of the target WASP-69 and the reference star in imaging mode of the GTC/OSIRIS. WASP-69 was placed in CCD 1 and the reference star was placed in CCD 2. \textit{Bottom panel}: Raw science image after activating spectroscopy mode.}
   \label{fig:fov}%
    \end{figure}

The observations were taken on October 5, 2016, using the OSIRIS spectrograph  on the GTC \citep{2000SPIE.4008..623C} in long-slit spectroscopy mode. For this, the planet-host star and a reference star of similar brightness  (see Table  \ref{tab:coord} for coordinates and magnitude information) were both placed in the slit (see Fig. \ref{fig:fov}). The slit was 40 arc seconds wide and custom-built for exoplanet atmosphere observations. The width of the slit allowed us to minimize possible flux losses in the event of seeing variability or small position drifts of the star within the slit during the observations. The observations were done using the R1000R grism, covering the wavelength range 510 - 905 (redder wavelength up to 1000 nm are covered but heavily affected by fringing and are thus discarded). The observations began at 20:23 UT, 49 min before the transit ingress; they lasted for 3 hr 36 min and ended at 00:00, 34 min after the transit egress. The observations were stopped prematurely due to a problem with the data acquisition system (DAS). The individual exposure times were set to 4 s, which yielded 591 spectra in total (367 during transit and 223 out of transit). At the beginning of the observations, the airmass was 1.6, and toward the end it was 1.23. The average seeing was 1.7 arcsec. A $2 \times 2$  binning was used, the gain was 1.46 $e^{-}$/ADU, the readout speed was 500 KHz, and the readout noise was 8.0 $e^{-}$.

\subsection{Data reduction}
The data were reduced following the approach described by \cite{2017A&A...600A.138C}, taking into account overscan, bias, flat field, and sky background correction. For the wavelength calibration, we made use of the full 2D information of the arc lamp exposures to create a pixel to wavelength transformation map. This assured the correct wavelength calibration for each obtained spectra, even if the spectral trace changed between exposures. The spectra were extracted along the spectral trace using the optimal extraction algorithm \citep{1986PASP...98..609H}. We tested several extraction apertures and compared the out-of-transit point to the point scatter of the light curve. The lowest scatter was found when we use a fixed aperture of 42 pixels for both target and reference stars, and we adopted this aperture for the following analysis. Time stamps for every spectrum were calculated to correspond to the middle of the exposure and converted into the Barycentric Julian Date (BJD) in the Barycentric Dynamical Time standard following  \cite{2010PASP..122..935E}. To create light curves for both stars, the flux in each spectrum was integrated over a given pass band. For the white light curve, the flux was integrated over the range 515 - 905 nm, excluding the region of low flux caused by the strong oxygen absorption band between 757 - 768 nm. The absolute light curve of the target star, WASP-69, is divided by the absolute light curve of the reference star thus creating a relative light curve. This division allow us to correct for effects of variable telluric absorption, seeing and atmospheric extinction over the course of the observation. In Fig. \ref{fig:whitelightcurve} we show the absolute light curves of both stars together with the relative transit light curve.

\begin{figure}
    \centering
    \includegraphics[width=9cm]{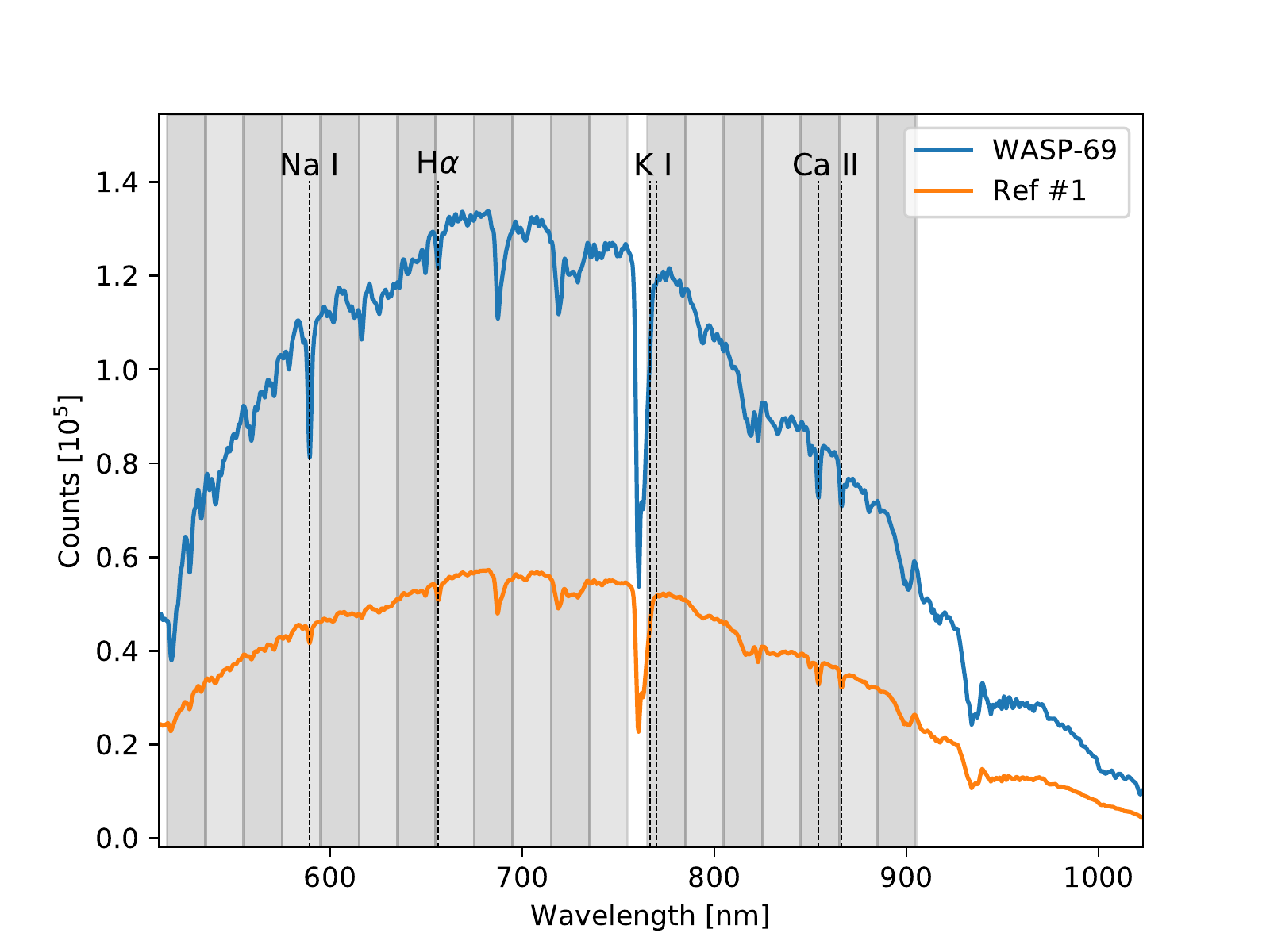}
    \caption{Sample spectra of WASP-69 (blue) and the reference star (orange). The limits of the narrow pass bands used to create the 19 spectrophotometric light curves are indicated in gray lines. The position of several strong atmospheric lines are also marked. The spectral region at~760 nm is excluded from narrow pass bands as it is strongly affected by telluric oxygen absorption, introducing noise to the data.}
    \label{fig:examplespec}%
  \end{figure}

To calculate relative light curves for 19 spectroscopic channels, the spectral range was divided into 20 nm wide bins. These bins are indicated in Fig. \ref{fig:examplespec} for spectra of both stars. For each 20 nm wide interval, the process described for the creation of a white light curve was repeated using a narrower wavelength interval. The resulting spectroscopic light curves are shown in Fig. \ref{fig:speclightcurves}.

\section{Data analysis}
\label{sec:analysis}
\subsection{White light curve}

In order to derive the wavelength independent transit parameters and fix them in the subsequent analysis of the spectroscopic light curve analysis, we fitted the white light curve (see \citealp{2018A&A...616A.145C} for details). For the fitting process, we used the python package \texttt{batman} \citep{2015PASP..127.1161K}, which describes the transit light curve using the analytical model by \cite{2002ApJ...580L.171M}. We accounted for systematic trends (red noise) in the light curves using Gaussian processes (GPs; \citealp{2006GP}, \citealp{2012MNRAS.419.2683G}, \citealp{2015ITPAM..38..252A}). The major sources of red noise that we considered in our modeling were the seeing variation (measured by the FWHM of the spatial profile), the position of the target star along the spatial direction, and a time component. These vectors were then used as input to a squared exponential kernel, with the amplitude and scale of the kernel set as free parameters.

The quadratic limb darkening coefficients $u_1$ and $u_2$ were calculated by interpolating a stellar atmosphere model with the stellar effective temperature of the host star ($T_{\rm eff}=4700$ K, surface gravity  log$g_{\star} =4.50,$ and metallicity [Fe/H]$= 0.150$ \citep{2014MNRAS.445.1114A}) from the ATLAS synthetic models \citep{1979ApJS...40....1K} using the code from \cite{2015MNRAS.450.1879E}.

\begin{figure}
   \centering
   \includegraphics[width=8cm]{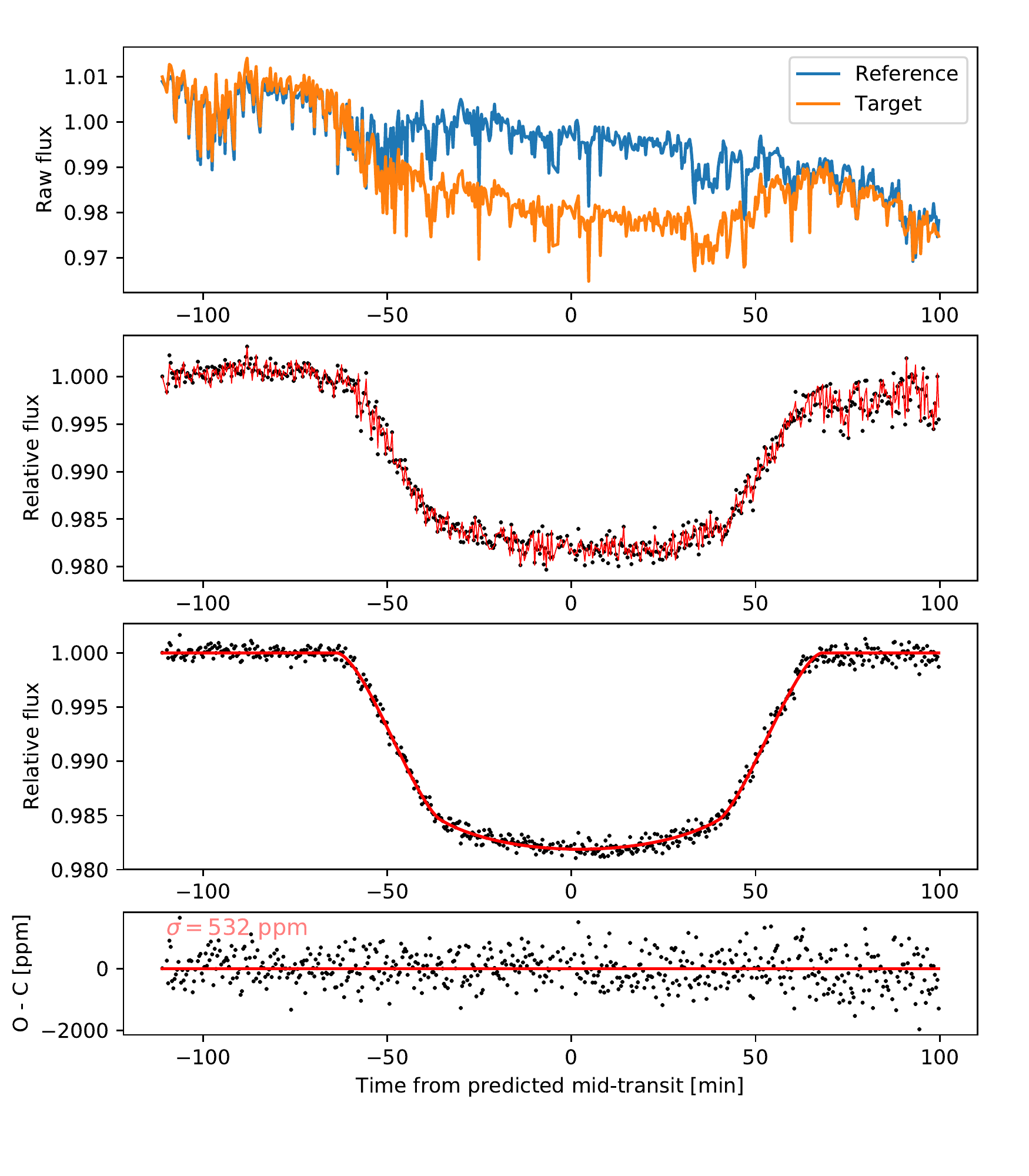}
   \caption{White light transit curve of WASP-69b. \textit{Upper panel}: raw light curve of WASP-69 (orange) and  the reference star (blue). \textit{Second panel from the top}: relative white light curve (black points) with best fitting model (in red), including the modeled transit curve and components to model the effects of seeing and other systematic noise sources. \textit{Third panel from the top}: relative white light curve with the modeled noise components removed. The best fitting transit model is plotted in red. \textit{Bottom panel}: residuals after the best fitting model is subtracted from the data are plotted in black with a red line indicating the zero level. The scatter of the residuals is 532 ppm.}
   \label{fig:whitelightcurve}%
 \end{figure}

In the transit light curve fit, the free parameters were the planet-to-star radius ratio $R_p/R_\star$, the semi major axis in units of the stellar radius $a_p/R_\star$, the orbital inclination $i$, and the mid-transit time $T_{\rm mid}$. The period and the orbital eccentricity were fixed to the values from the discovery paper \citep{2014MNRAS.445.1114A} ($e=0$). The MCMC procedure consisted of 90 chains, each one with 3000 iterations. The final values and associated errors of the fitted parameters were estimated using the percentiles of the posterior distribution (median and 1-$\sigma$ uncertainties). The best fitting transit parameters for the white light curve fit are given in Table \ref{tab:wlresult}. The best fitting models are shown together with the data in Fig. \ref{fig:whitelightcurve}.

\begin{table} 
\caption{Best-fitting planet system parameters from optimization using GPs for the white light curve of WASP-69b.}

\begin{center}
\def\tol#1#2#3{\hbox{\rule{0pt}{15pt}${#1}^{{#2}}_{{#3}}$}}
\begin{tabular}{ll}
\hline\hline 

Transit parameter &Value \\  \hline

$R_\mathrm{p}/R_{\star}$ & \tol{0.13238}{+0.00091}{-0.00094} \\
$a_\mathrm{p}/R_{\star}$ & \tol{12.07}{+0.203}{-0.198} \\
$T_\mathrm{mid}[\mathrm{BJD}_{\mathrm{TBD}}] $ & 2457667.432296(171) \\
Period [d] & 3.8681382\tablefootmark{a} (fixed) \\
$i$ (deg) & \tol{86.692}{+0.122}{-0.123} \\
$\omega$ (deg) & 90\tablefootmark{a} (fixed)\\
$e$ & 0\tablefootmark{a} (fixed)\\
$u_1$ & \tol{0.451}{+0.067}{-0.067} \\
$u_2$ & \tol{0.118}{+0.083}{-0.084} \\

\hline
\end{tabular}
\tablefoot{
\tablefoottext{a}{\citet{2014MNRAS.445.1114A}.}}
  \label{tab:wlresult}
\end{center}
\end{table}

\subsection{Spectroscopic light curves}
\begin{figure*}
    \centering
    \includegraphics[width=18cm]{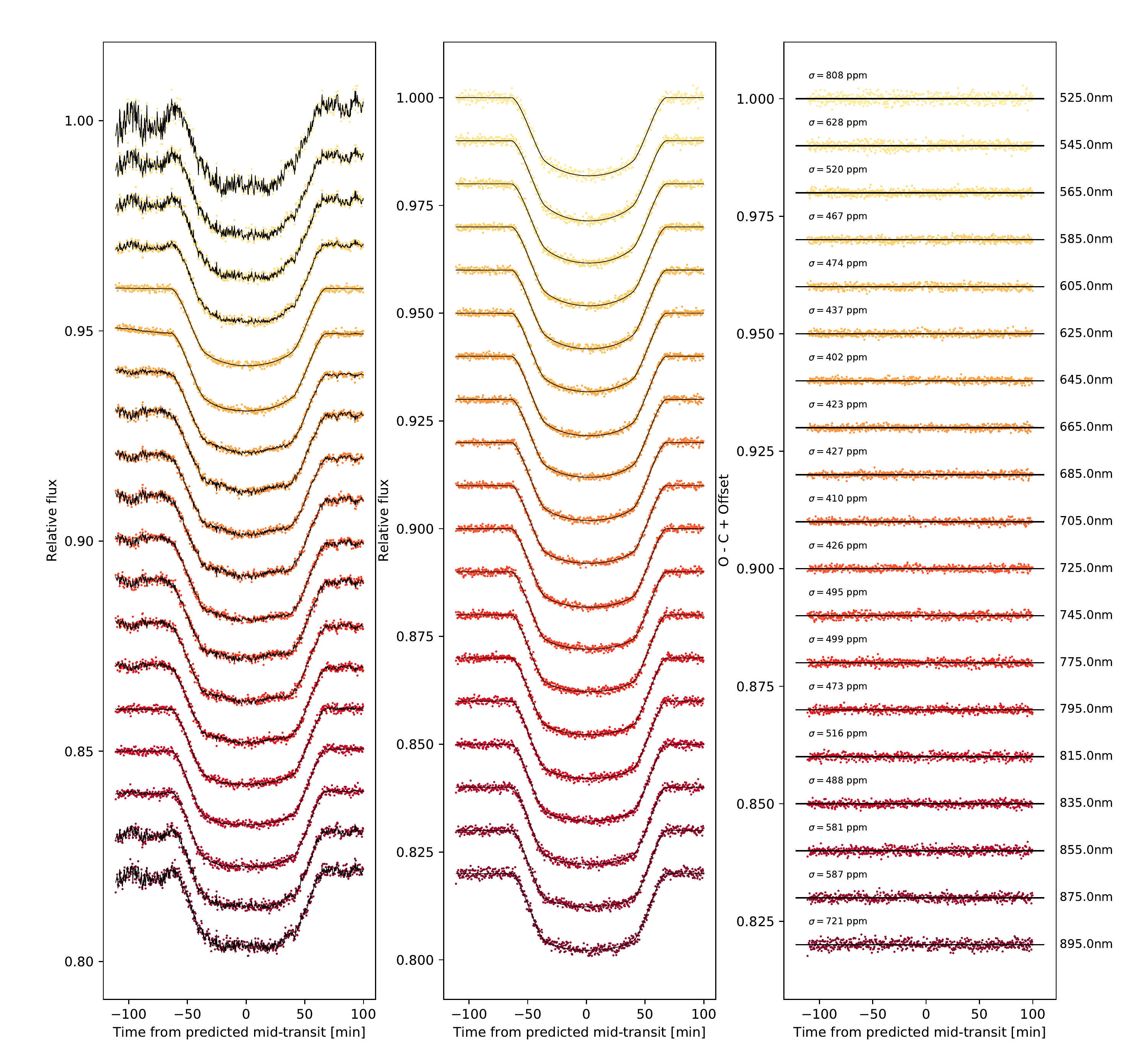}
    \caption{\textit{Left panel}: all 19 spectroscopic light curves where common-mode noise has been removed, with the respective best fitting model (including transit light curve and noise model) plotted in black. \textit{Middle panel}: same spectroscopic light curves after the model for systematic trends was removed. The best fitting transit model considering the red noise is plotted in black. \textit{Right panel}: residuals after the best fitting models were subtracted. }
    \label{fig:speclightcurves}%
  \end{figure*}

For the analysis of the spectroscopic light curves, the wavelength independent transit parameters were fixed to the values obtained for the white light curve, and only the planet-to-star radius ratio and the limb darkening coefficients were allowed to vary. We created a common-mode noise model by dividing the white light curve by the best analytical transit model fit, then we divided each individual light curve by this common-mode noise model. Next, we fitted a transit model together with a red noise model using the same inputs as the white light curve (i.e., a squared exponential kernel with seeing, star position in spatial direction, and time component as inputs) to the spectroscopic light curves corrected by the common-mode noise model. To estimate the best-fitted parameter values and their 1-$\sigma$ uncertainties, we used 32 chains running for 3000 iterations each; after that, we obtained the fitted values and uncertainties from the posterior distribution of the parameters. In Fig. \ref{fig:speclightcurves} all 19 curves are shown together with the best fitting model, including the GP noise model.

\section{GTC/OSIRIS optical transmission spectrum}
\label{sec:TransOptical}

\begin{figure*}
   \centering
   \includegraphics[width=15cm]{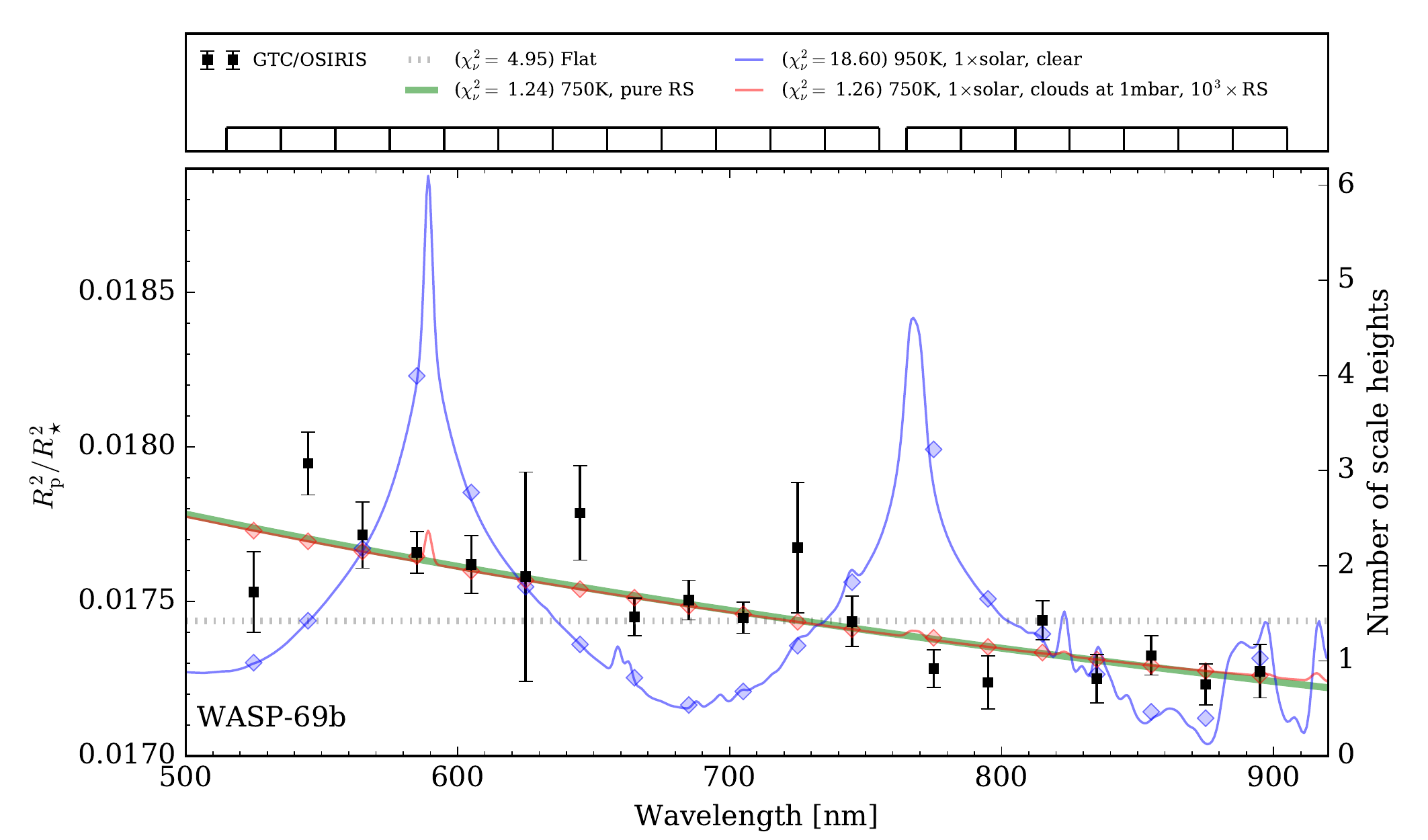}
   \caption{Optical transmission spectrum of WASP-69b (black squares) compared with several theoretical atmosphere models. In blue, a model created using \texttt{Exo-Transmit} (\citealp{2017PASP..129d4402K}) for a clear atmosphere and solar abundance is shown. The gray dotted line indicates a model of flat transmission spectrum. The slope expected in an atmosphere affected solely by Rayleigh scattering is shown in green. In red, we show the model of an atmosphere with clouds at 1 mbar and Rayleigh scattering enhanced by a factor of 1000. {\it Top panel} shows the legend and the bins of bandpass to derive the transmission spectrum.}
   \label{fig:transspecopt}%
   \end{figure*}

 The resulting wavelength dependent planet-to-star radius ratio is given in Table \ref{tab:specrprs}, and the resulting optical transmission spectrum of the exoplanet is plotted in Fig. \ref{fig:transspecopt}. In that figure, the transmission spectrum is compared to two different synthetic models, as well as to a Rayleigh scattering slope and to a line indicating a completely flat atmospheric transmission.

The data show no indication of absorption by sodium or potassium, ruling out the model of a clear atmosphere with solar abundance at high confidence. Comparison of the data to a flat line also rejects a cloudy, flat transmission spectrum. Instead, the data exhibit a clear trend of increasing planet radius toward shorter wavelengths. This slope is compatible with the expected increase due to Rayleigh scattering.

\subsection{Rayleigh scattering}
The equation that predicts the change of the measured planetary radius across wavelength in a Rayleigh scattering regime is (\citealp{2008A&A...481L..83L}):

\begin{align}
\frac{\rm{d}R_{\rm{p}}}{\rm{d}\ln{\lambda}}=\alpha H=\alpha \frac{\kappa_{\rm{B}} T}{\mu_{\rm{m}} g_{\rm{p}}},
\end{align}
where $H$ is the atmospheric scale height, $\kappa_{\rm{B}}$ is the Boltzmann constant, $T$ is the planet's atmospheric temperature, $\mu_{\rm{m}}$ is the mean molecular weight of the planetary atmosphere, and $g_{\rm{p}}$ is the surface gravity of the planet. If the particles that are causing the scattering are mainly hydrogen, it is expected that $\alpha = -4.0$.

Assuming a hydrogen-dominated atmosphere ($\mu_{\rm{m}}=2.37$), an atmospheric temperature equal to the planetary equilibrium temperature ($T_{\rm{eq}}=963 \pm 18$ K), and $\log g_{\rm{p}} = 2.726 \pm 0.046$ (cgs) (values taken from \citealp{2014MNRAS.445.1114A}), we find $\alpha = -3.35 \pm 0.75$, a value consistent with the expected metrics from hydrogen. Another piece of indirect evidence that support the hypothesis that the observed slope is caused by hydrogen is the escaping tail of helium discovered by \cite{2018Sci...362.1388N}. If helium is detected in the upper atmosphere of WASP-69b, it is likely that there is also a detectable amount of hydrogen in its atmosphere. However, since WASP-69 is an active star with evidence of strong emission in Ca {\sc ii} H+K lines ($\log R'_{HK} \sim -4.54$, \citealp{2014MNRAS.445.1114A}), it is possible that the slope that we detect with the GTC could also be caused by stellar activity.

\subsection{Stellar spots and faculae}
\label{sec:spot}

As WASP-69b is an active star, the role of occulted and unocculted stellar spots and faculae, and their ability to imprint themselves on the transmission spectrum, need to be taken into account. 
We do not detect any obvious deformation in the light curves indicative of stellar spots occulted by the planet during the observed transit. However, a slope toward larger radii with short wavelengths can be caused by unocculted spots on the stellar disk (e.g., \citealp{2013MNRAS.432.2917P}, \citealp{2014ApJ...791...55M}, \citealp{2014A&A...568A..99O}). 
Unocculted spots can cause an increase in the apparent planet-to-star radius ratio toward the blue; this is due to the wavelength dependent contrast between a cold spot and the remaining stellar surface, which increases toward shorter wavelengths. On the other hand, faculae (zones with slightly hotter temperatures than the average of the star) can cause the opposite effect: an increase of the apparent planet-to-star radius ratio toward the redder parts of the optical spectrum. A simple way to describe the expected change in apparent radius ratio is given by Eq. \ref{Eq:RpRs_spots} (\citealp{2018ApJ...853..122R} and references therein); here, $\delta_{spot}$ and $\delta_{facu}$ are the area ratio between the spot- and faculae-covered area and the entire stellar disk, $F_{\nu} \left(\rm{spot}\right)$ and $F_{\nu} \left(\rm{facu}\right)$ are the fluxes from the spots and faculae, and $F_{\nu} \left(\rm{phot}\right)$ is the flux of the stellar disk:

\begin{equation}
   \left(\frac{\hat{R}_{\rm{p}}}{\hat{R}_{\star}}\right)^2= \left(\frac{R_{\rm{p}}}{R_{\star}}\right)^2
   \frac{1}{1- \delta_{spot}\left(1- \frac{F_{\nu} \left(\rm{spot}\right)}{F_\nu \left(\rm{phot}\right)}\right)- \delta_{facu}\left(1- \frac{F_{\nu} \left(\rm{facu}\right)}{F_\nu \left(\rm{phot}\right)}\right)}
   \label{Eq:RpRs_spots}
.\end{equation}

We explore this scenario by fitting the transmission spectra using a grid of PHOENIX stellar spectra (\citealp{2013A&A...553A...6H}) with different temperatures (ranging from 2700 K to 7000 K) but with the same metallicity and $\log g_{\star}$ as WASP-69 (taken from \citealp{2014MNRAS.445.1114A}) to model the temperature of the star, the spots, and the faculae. As described in \cite{2019A&A...622A.172M}, an MCMC fitting was done using \texttt{emcee} (\citealp{2013PASP..125..306F}); we used 80 chains and a two-step iteration procedure (7500 iterations as the burn-in phase and 25000 iterations as the main MCMC run). The type of prior and range of values used for each parameter is presented in Table \ref{tab:SpotMCMC}. After the MCMC fit was done, the best fitted values were adopted from the posterior distributions of the fitted parameters (the median from the distribution as the final parameter value and 1$\sigma$ uncertainty range as the respective error bar).

\begin{table}
 \caption[]{Spot modeling fit.}
\begin{center}
\def\tol#1#2#3{\hbox{\rule{0pt}{15pt}${#1}^{{#2}}_{{#3}}$}}
\setlength{\tabcolsep}{1.5mm}
\begin{tabular}{c l c}
\hline\hline
  Parameter &  Prior type and range &  Fitted value \\\hline  
  $\left(R_{\rm{p}}/R_{\star}\right)$ &  Uniform [0.06,0.26] & \tol{0.12880}{+0.00042}{-0.00041} \\
  $\delta_{\rm{spot}}$ & Uniform [0.0,1.0] & \tol{0.553}{+0.301}{-0.267} \\
  $\delta_{\rm{facu}}$ & Uniform [0.0,1.0] & \tol{0.149}{+0.460}{-0.132}  \\
  $T_{\rm{phot}}$ & Normal [$\mu=4700$,$\sigma=50$] & \tol{4716}{+22.6}{-33.4} K\\
  $T_{\rm{spot}}$ & Uniform [2700,4700] & \tol{4594}{+48.1}{-77.4} K \\
  $T_{\rm{facu}}$ & Uniform [4700,7000] & \tol{4788}{+307.8}{-68.2} K\\
  \hline
\end{tabular}
\label{tab:SpotMCMC}
\end{center}
\end{table}

The results of the best fitting spot and faculae coverage model that is able to reproduce our observed transmission spectrum are presented in Table \ref{tab:SpotMCMC}, and Fig. \ref{FigSpotCorner} presents the correlation plots of the fitted parameters. The model converges to a star temperature of $T_{\rm{phot}} = 4716^{+23}_{-33}$ K, a spot temperature of $T_{\rm{spot}} = 4594^{+48}_{-77}$ K, and a faculae temperature of $T_{\rm{facu}} = 4788^{+308}_{-68}$ K. The temperature contrast between the stellar surface and the spots is $\Delta T_{\rm{phot-spot}} = +122$ K, and the temperature contrast stellar surface versus faculae is $\Delta T_{\rm{phot-facu}} = -72$ K. The filling factors for the spots and faculae are $\delta_{\rm{spot}} = 0.55^{+0.30}_{-0.27}$ and $\delta_{\rm{facu}} = 0.15^{+0.46}_{-0.13}$, respectively. As mentioned in \cite{2019A&A...622A.172M} and references therein, spot coverage on the order of 40-50\% for active K dwarfs are not rare; however, the correlation plot of the fit shows that the filling factors of the spots and faculae are not well constrained and are degenerate with the temperature of their respective features.

A combination of stellar features, such as spots and faculae, can explain the slope we detect with the GTC, which would make the transmission spectrum of WASP-69b similar to that of the Neptune-sized planet HAT-P-11b. In one GTC observation, HAT-P-11b presented  a Rayleigh-like feature, which was attributed to spots, in contrast to the flat transmission spectrum observed on another occasion with the same instrument (\citealp{2019A&A...622A.172M}). However, unlike HAT-P-11b, for WASP-69b there are no reported cases of spot regions big enough to be detected as spot-crossing events in the follow-up observations made in the discovery paper by \cite{2014MNRAS.445.1114A}, although there is evidence of stellar modulation due to spots in their long trend light curves. Nor do we detect any spot-crossing in our high-precision white-light photometric light curve. Future monitoring of several WASP-69b transits may help resolve this issue.

\subsection{Exploring the Na {\sc i} doublet region for planetary absorption}

Using HARPS-North high-resolution spectroscopy ($R\sim 115000$), \cite{2017A&A...608A.135C} reported an excess of the measured planetary radius of WASP-69b around the Na {\sc i} doublet (absorption lines D2 with $\lambda_{\mathrm{D2}} = 589.0$ nm and D1 $\lambda_{\mathrm{D1}} = 589.6$ nm ). In contrast, \cite{2019AJ....157...58D} did not find the excess of the planetary radius around the Na {\sc i} lines in their study of Warm Saturns using the ESPaDOnS ($R \sim 60000$) instrument on CFHT. Although low-resolution transmission spectroscopy is not particularly sensitive to the core of absorption lines, it can still be a useful tool to detect Na in transiting planets by testing the detection of broad line wings in clear atmospheres.

 We explored the region around the Na {\sc i} doublet using 11 spectroscopic channels with 5 nm widths to see if we were able to detect an excess of the measured planet-to-star radius ratio that would confirm the presence of Na in the atmosphere of WASP-69b. Table \ref{tab:specrprsNa} presents the measured $R_{\rm{p}}/R_\star$ with their 1$\sigma$ uncertainties, and Fig. \ref{fig:Naline} shows the transmission spectrum around the Na {\sc i} doublet and compares the measurements with different models. We detected neither an extra absorption in the bin centered at the Na {\sc i} doublet, nor the broad absorption line wings associated with a clear atmosphere (i.e., an atmosphere without clouds or hazes that mute the Na absorption signal). {\cite{2017A&A...608A.135C} measured a line contrast of 5.8\% for the D2 line and had no detection of D1. The average seeing of 1.9$''$ corresponds to $\sim$20 $\AA$ for the OSIRIS observation. If we integrate the high-resolution excess absorption line profile and convolve it with the seeing-limited spectral resolution, we measure a transit depth difference of 44~ppm in a bin width of 5 nm. This value is much smaller than our typical uncertainty of $\sim$120~ppm around the Na doublet (see Table \ref{tab:specrprsNa}), hence we would not be able to detect the Na absorption presented in \cite{2017A&A...608A.135C} with our GTC low-resolution data.} 

As mentioned in \cite{2019A&A...622A.172M}, the measured planet-to-star radius ratio centered around the Na {\sc i} doublet can be affected by unocculted stellar spots (for a more detailed discussion, see \citealp{2019AJ....157...96R}). In Fig. \ref{fig:Naline} the best fitted model that takes into account unocculted spots and faculae is shown in orange. The predicted amplitude of the Na feature produced by unocculted spots is significantly smaller than the amplitude predicted by planetary atmospheric models and is of comparable size to the uncertainty of our measurements.

   \begin{figure*}
   \centering
   \includegraphics[width=15cm]{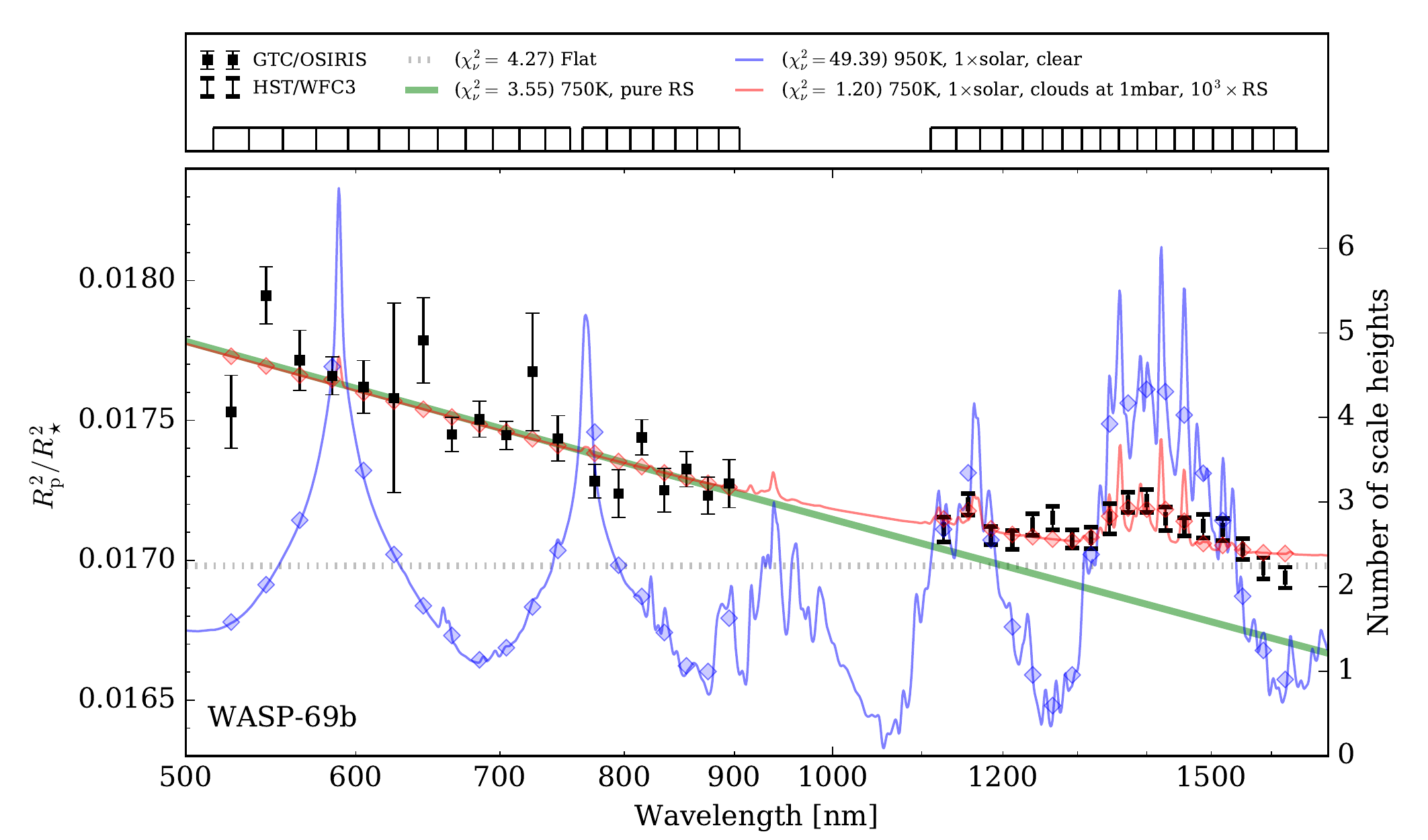}
   \caption{Optical and near-infrared transmission spectrum of WASP-69b (black squares) compared with the same theoretical atmosphere models explained in Fig. \ref{fig:transspecopt}. The near-infrared data was obtained using WFC3 \citep{2018AJ....155..156T}. The WFC3 data have been shifted upward by a constant value of $\Delta {R_{\rm{p}}^2/R_\star^2} = 580$~ppm.} 
   \label{fig:FigGTCHST}%
   \end{figure*}

  \begin{figure*}
    \centering
    \includegraphics[width=15cm]{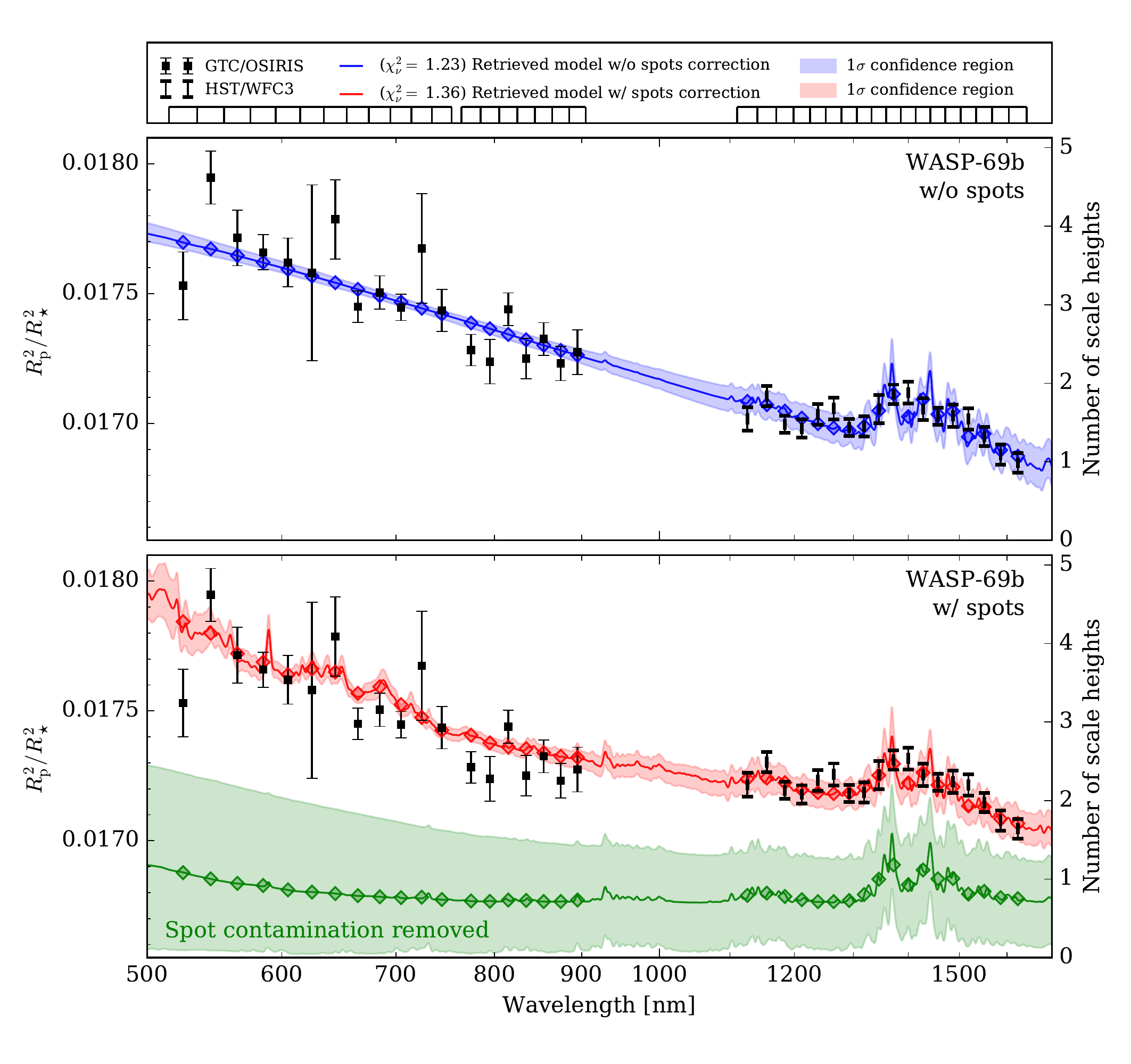}
    \caption{Optical and near-infrared transmission spectrum of WASP-69b (black squares) compared with the retrieved models from the \texttt{PLATON} code \citep{2019PASP..131c4501Z}. The shaded area presents the 1$\sigma$ confidence level of the retrieval analysis. {\it Top panel} shows the legend and the bins of bandpass used to derive the transmission spectrum.{\it Middle panel} shows the retrieval analysis assuming no spot contamination. {\it Bottom panel} shows the retrieval analysis with spot contamination, where the green model corresponds to the pure planetary atmospheric model after removing the spot contamination from the red model. The WFC3 data \citep{2018AJ....155..156T} have been shifted upward by a constant value of $\Delta {R_{\rm{p}}^2/R_\star^2} = 479$~ppm (middle panel) and 618~ppm (bottom panel), respectively, which is a free parameter in the retrieval analysis.}
    \label{fig:retrieved}
  \end{figure*}

\section{Optical and near-infrared transmission spectrum of WASP-69b}
\label{sec:TransFull}

We put our transmission spectrum observed in the optical range into context by comparing it to the results obtained by \cite{2018AJ....155..156T} in the near infrared using WFC3 at the HST, as shown in Figs. \ref{fig:FigGTCHST} and \ref{fig:retrieved}. Their results exhibit the detection of a muted water band together with a sloped continuum. 

The planet ratio, derived from the OSIRIS transmission spectrum, is higher than that from the WFC3 transmission spectrum with an overall offset of $\Delta R_\mathrm{p}/R_\star\sim 0.00221$. \citet{2018A&A...620A.142A} find that different sets of orbital parameters (e.g., $a/R_\star$ and $i$) could systematically introduce discrepancy to the transmission spectrum. In our case, if we fixed the transit parameters to the ones adopted by WFC3 (i.e., $a/R_\star=11.953$ and $i=86.71^{\circ}$), the OSIRIS transmission spectrum would hold the same spectral shape but move downward with an overall offset of $\Delta R_\mathrm{p}/R_\star=0.00076$, which cannot account for the observed offset of $\Delta R_\mathrm{p}/R_\star\sim 0.00221$. The remaining offset could come from the bias introduced by different instrumental systematics or different flux levels caused by the modulation of stellar activity. If the remaining offset is solely introduced by stellar activity, it would indicate that the flux level at the epoch of the WFC3 observation is $\sim$2.2\% higher than that at the epoch of the OSIRIS observation. The discovery WASP photometry of WASP-69 indeed shows several modulation periods with different amplitudes. The largest semi-amplitude of $\sim$13~mmag would allow a flux variation of $\sim$2.4\%, on par with the aforementioned remaining offset.

In the following sections, we compare the observed spectra with theoretical models to infer the potential atmospheric properties of WASP-69b. 

\subsection{Forward model comparison}

We created a grid of transmission spectrum fiducial models using the \texttt{Exo-Transmit} (\citealp{2017PASP..129d4402K}) code. The models adopted isothermal temperature profiles spanning from 450~K to 1450~K in steps of 100~K. The adopted metallicities were 0.1$\times$, 1$\times$, 10$\times$, 100$\times$, and 1000$\times$ solar. We also considered cloud-free scenarios and clouds at 10, 1, and 0.1~mbar, along with scattering of amplitudes at 1$\times$, 10$\times$, 100$\times$, 1000$\times$, and 10000$\times$ H$_2$ Rayleigh scattering. We binned the transmission spectrum models into the pass bands of the OSIRIS and WFC3 transmission spectra. We compared the observed data to the binned models, allowing the WFC3 data to have an overall offset as the free parameter. 

The best-matched model comes from the atmosphere with a temperature of 750~K, a metallicity of 1$\times$solar, a 1000$\times$ enhanced Rayleigh scattering, and clouds at 1~mbar. Figure \ref{fig:FigGTCHST} shows the observed data compared to this best-matched model ($\chi^2_{\mathrm{r}}= 1.20$), together with another three models: i) a flat line ($\chi^2_{\mathrm{r}}= 4.27$); ii) a pure Rayleigh scattering model ($\chi^2_{\mathrm{r}}= 3.55$); and iii) a cloud-free model with 1$\times$solar metallicity at the equilibrium temperature of $\sim$950~K ($\chi^2_{\mathrm{r}}= 49.39$). The completely cloud-free and the completely cloudy atmosphere models can be ruled out. This indicates that an atmosphere with clouds and enhanced Rayleigh scattering from hazes can explain both the optical and near-infrared results.

\subsection{Spectral retrieval analysis}
  
\begin{table}
\caption{Retrieved atmospheric parameters using \texttt{PLATON} models with only atmospheric scattering (i.e., without stellar spots) and for models with atmospheric scattering plus the effect of stellar spots.}
\begin{center}
\def\tol#1#2#3{\hbox{\rule{0pt}{15pt}${#1}^{{#2}}_{{#3}}$}}
\setlength{\tabcolsep}{1.5mm}
{
\begin{tabular}{l c c}
\hline\hline
Parameter &  w/o spots  &  w/ spots  \\\hline
Planet radius at 1~bar $R_\mathrm{p}^\prime$ [$R_\mathrm{p}$] & \tol{0.947}{+0.006}{-0.008} & \tol{0.945}{+0.007}{-0.017} \\
Temperature $T_\mathrm{p}$ [K] & \tol{1227}{+133}{-164} & \tol{1203}{+123}{-429} \\
Scattering slope $\alpha$ & \tol{-8.9}{+2.1}{-2.3}  & \tol{-8.3}{+2.6}{-6.7} \\
Scattering amplitude $\log A$ & \tol{5.2}{+0.3}{-0.4} & \tol{2.4}{+1.4}{-3.6} \\
C/O ratio & \tol{0.37}{+0.19}{-0.20} & \tol{0.41}{+0.19}{-0.20} \\
Metallicity $\log Z/Z_\sun$ & \tol{2.7}{+0.1}{-0.2} & \tol{2.7}{+0.2}{-0.4} \\
Cloud-top pressure $\log P_\mathrm{top}$ [Pa] & \tol{2.8}{+1.4}{-1.7} & \tol{0.5}{+0.8}{-0.4} \\
Spot temperature $T_\mathrm{spot}$ [K] & -- & \tol{4244}{+236}{-171} \\
Spot fraction $f_\mathrm{spot}$ & -- & \tol{0.10}{+0.03}{-0.01} \\
WFC3 data offset $\delta_\mathrm{WFC3}$ [ppm] & \tol{-479}{+55}{-58} & \tol{-618}{+60}{-47}  \\
        \hline
\end{tabular}
}
\label{tab:retrieved}  
\end{center}
\end{table}

We performed spectral retrieval analysis on the combined OSIRIS+WFC3 transmission spectra using the \texttt{PLATON} \citep{2019PASP..131c4501Z} code. The latest version, 5.0, allows the WFC3 data to have an overall shift as the free parameter. We conducted two runs of retrievals: one assuming atmospheric scattering and no contamination by stellar spots, and one that included scattering and the effect of spots. The free parameters include: planet radius at 1~bar ($R_p^\prime$), atmospheric temperature $T_p$, C/O ratio, atmospheric metallicity ($\log Z/Z_\sun$), cloud-top pressure ($\log P_\mathrm{top}$), scattering slope $\alpha$, and scattering amplitude $\log A$. In the case of spot correction, the spots are assumed to have a temperature of $T_\mathrm{spot}$ with a filling factor of $f_\mathrm{spot}$, which assume only one net effect (i.e., spots or faculae); this differs from the separated treatment in Eq. \ref{Eq:RpRs_spots}. We adopted uniform priors for these parameters and employed the nested sampling, implemented by \texttt{dynesty} \citep{2020MNRAS.493.3132S} with 1000 live points, to explore the parameter space. The retrieved results are presented in Table \ref{tab:retrieved} and Fig. \ref{fig:retrieved}.

Both retrieval runs can fit the observed transmission spectrum reasonably well, and both yield a temperature of $\sim 200$ K higher than the planetary equilibrium temperature ($T_{\mathrm{eq}}= 963 \pm 18$ K, \citealp{2014MNRAS.445.1114A}). The retrieved metallicity tends to be super-solar, and the retrieved scattering slope is steeper than the nominal Rayleigh scattering slope ($\alpha=-4$). The run with spot correction delivers a spot temperature of $\Delta T_{\rm{phot-spot}} = 456^{+236}_{-171}$~K lower than the stellar photosphere, with a spot coverage of $0.10^{+0.03}_{-0.01}$; that is to say, a colder spot temperature and lower spot coverage factor than the values obtained using only the optical data (see Sect. \ref{sec:spot}).

\subsection{The atmosphere of WASP-69b}

If we assume a molecular weight of $\mu_{\mathrm{m}} = 2.37$, our GTC/OSIRIS optical results are consistent with Rayleigh scattering caused by the presence of hydrogen in the atmosphere of WASP-69b. Since WASP-69 is an active star, it is also feasible that the observed slope is solely caused by stellar activity in the form of spots and faculae. If unocculted spots are causing the slope seen in the optical transmission spectrum, there could be a detectable extra absorption at the Na {\sc i} doublet; however, according to our spot and faculae modeling, that extra absorption is on the order of our measured planet-to-star radius ratio uncertainties.

Combining our optical results with a previously published HST/WFC3 near-infrared transmission spectrum, we made two modeling tests: forward model comparison and spectral retrieval. With forward modeling comparison, we can rule out the completely cloud-free or completely cloudy models. Spectral retrieval (with and without considering the effect of stellar spots) seems to suggest that extra sources for scattering (i.e., hazes) and super-solar metallicity can explain the observed slope (and atmospheric features in general) in both wavelength regimes. Hence, there is a possible physical solution that indicates that the detected Rayleigh scattering signal can come from the planet's atmosphere. 
Both derived transmission spectra (optical and near-infrared) present a slope compatible with Rayleigh scattering; if this slope is caused by stellar activity in the form of unocculted spots, it would mean that the spot coverage was more or less similar during both observations. The HST/WFC3 data were taken on August 16, 2016 (Proposal ID 14260, P. I. Drake Deming), while the GTC/OSIRIS data were taken on October 5, 2016, roughly 50 days apart, or close to twice the rotational period of WASP-69 ($P_{\mathrm{rot}} = 23$ days, \citealp{2014MNRAS.445.1114A}). Studies performed on other active planet-host stars found spot life cycles between 60 and 90 days (Kepler-210, a K-type star, \citealp{2016A&A...594A..41I}) and even evidence for longer lifetimes (75 to 330 days, Kepler-17, G2V star, \citealp{2012A&A...547A..37B}); the possibility that both observations were affected by the same (or a similar) group of spots present on the surface of WASP-69 cannot be ruled out.

The optical and near-infrared transmission spectrum of HAT-P-12b was announced by \cite{2020AJ....159..234W} as we were writing this paper. HAT-P-12b is a sub-Saturn mass transiting planet orbiting a K dwarf discovered by \cite{2009ApJ...706..785H}. This planet possesses a mass, radius, and orbital period ($M_\mathrm{p} = 0.21 \; M_{\mathrm{Jup}}$, $R_\mathrm{p} = 0.96 \; R_{\mathrm{Jup}}$, and $P = 3.21$ days) similar to those of WASP-69b ($M_\mathrm{p} = 0.26 \; M_{\mathrm{Jup}}$, $R_\mathrm{p} = 1.05 \; R_{\mathrm{Jup}}$, and $P = 3.86$ days); however, HAT-P-12b orbits around a star with sub-solar metallicity ($[\mathrm{Fe/H}] = -0.29$ dex) and WASP-69 metallicity is super-solar ($[\mathrm{Fe/H}] = 0.15$ dex). The optical and near-infrared transmission spectrum of HAT-P-12b looks remarkably similar to that of WASP-69b, with a slope consistent with Rayleigh scattering in the optical and a water absorption feature at 1.4 $\mu$m. Based on their photometric follow-up, \cite{2020AJ....159..234W} argue that HAT-P-12 is a very quiescent star with low photometric variability and conclude that the observed optical slope in the spectrum likely originated from the planetary atmosphere. In contrast, WASP-69 seems to exhibit higher levels of stellar activity, as indicated by the $\log R'_{HK}$ index ($\log R'_{HK} \sim -4.54$, \citealp{2014MNRAS.445.1114A}, compared to HAT-P-12 $\log R'_{HK} = -4.90,$ \citealp{2018A&A...613A..41M}) and stellar activity in the form of photometric variability, as shown in the discovery paper of this system. Nevertheless, their retrieved transmission spectra look identical.

\section{Conclusions}
\label{sec:conclusions}

We present here the results of the analysis of one primary transit of the Saturn-mass exoplanet WASP-69b, taken on October 5, 2016, with the long-slit spectrograph OSIRIS mounted on the GTC (10.4 m). We observed WASP-69 and one reference star simultaneously over $\sim 3.5$ hours. Using their spectra, we were able to integrate the flux of both stars, and we created several light curves using differential spectrophotometry.

We searched for evidence of Na absorption originating from the planetary atmosphere of WASP-69b that created light curves with central bands centered around the Na {\sc i} doublet. The transmission spectrum around the Na lines is mostly flat, with no signs of extra absorption at the center of the Na doublet and with no signs of broad absorption line wings. However, the expected Na I features reported in previous studies are probably below our detection limits. 

The observed optical transmission spectrum presents a slope of increasing planet-to-star radius ratio toward blue wavelengths. Using the equation that links the change of apparent planetary radius versus wavelength and atmospheric scale height, we find $\alpha = -3.35 \pm 0.75$, assuming that the atmosphere of WASP-69b is hydrogen-dominated ($\mu_{\rm{m}}=2.37$) and adopting a planetary temperature of $T_{\rm{eq}}=963 \pm 18$ K and a planetary surface gravity of $\log g_{\rm{p}} = 2.726 \pm 0.046$ (cgs). This value is consistent with Rayleigh scattering produced by hydrogen. Since WASP-69 is an active star, it is possible that the observed slope is caused by unocculted spots and faculae. Modeling the amount of area covered and the temperatures of spots and faculae necessary to reproduced our optical data, we find filling factors of $\delta_{\rm{spot}} = 0.55^{+0.30}_{-0.27}$ and $\delta_{\rm{facu}} = 0.15^{+0.46}_{-0.13}$ and temperatures $T_{\rm{spot}} = 4594^{+48}_{-77}$ K and $T_{\rm{facu}} = 4788^{+308}_{-68}$ K for spots and faculae, respectively.

We combined our optical GTC/OSIRIS measurements with a previously published HST/WFC3 near-infrared transmission spectrum. We compared the full set (optical plus near-infrared spectrum) with different atmospheric models; we can completely rule out cloud-free or completely cloudy models to explain the observations. Our spectral retrieval analysis (including the effect of stellar spots) suggests that extra sources for scattering and super-solar metallicity models can explain the main optical and near-infrared features of the transmission spectrum of WASP-69b.

Based on the previously reported slope found in the HST/WFC3 transmission spectrum of WASP-69b, on the previously reported atmospheric escape of He, on the lack of any spot-crossing features in our white-light transit light curve, and on our atmospheric model comparison, we speculate that the slope seen in our data is more likely to be caused by hazes present in the atmosphere of WASP-69b rather than by stellar active regions. However, we cannot completely rule out the possibility that the detected signal is indeed produced by stellar activity, hence independent observations of the system with other instruments at different epochs should help to confirm or reject this interpretation.


\begin{acknowledgements}
     Based  on  observations  made  with  the  Gran  Telescopio Canarias (GTC), installed in the Spanish Observatorio del Roque de los Muchachos  of  the  Instituto  de  Astrof\'{i}sica  de  Canarias,  in  the  island  of  La  Palma. G.C. acknowledges the support by the B-type Strategic Priority Program of the Chinese Academy of Sciences (Grant No. XDB41000000) and the Natural Science Foundation of Jiangsu Province (Grant No. BK20190110).\\
     \textit{Software}: \texttt{ipython} (\citealp{PER-GRA:2007}), \texttt{numpy} (\citealp{vanderwalt2011}), \texttt{scipy} (\citealp{Jones2001}), \texttt{matplotlib} (\citealp{Hunter2007}). This research made use of Astropy, a community-developed core Python package for Astronomy (\citealp{Astropy2013}, \citealp{Astropy2018}). Correlation plots for spot modeling done with \texttt{Corner} (\citealp{corner}).
\end{acknowledgements}

%
 
\bibliographystyle{aa.bst} 
\bibliography{GUESSw69.bib}

\begin{thebibliography}{67}
\expandafter\ifx\csname natexlab\endcsname\relax\def\natexlab#1{#1}\fi

\bibitem[{{Alexoudi} {et~al.}(2018){Alexoudi}, {Mallonn}, {von Essen},
  {Turner}, {Keles}, {Southworth}, {Mancini}, {Ciceri}, {Granzer}, {Denker},
  {Dineva}, \& {Strassmeier}}]{2018A&A...620A.142A}
{Alexoudi}, X., {Mallonn}, M., {von Essen}, C., {et~al.} 2018, \aap, 620, A142

\bibitem[{{Ambikasaran} {et~al.}(2015){Ambikasaran}, {Foreman-Mackey},
  {Greengard}, {Hogg}, \& {O'Neil}}]{2015ITPAM..38..252A}
{Ambikasaran}, S., {Foreman-Mackey}, D., {Greengard}, L., {Hogg}, D.~W., \&
  {O'Neil}, M. 2015, IEEE Transactions on Pattern Analysis and Machine
  Intelligence, 38, 252

\bibitem[{{Anderson} {et~al.}(2014){Anderson}, {Collier Cameron}, {Delrez},
  {Doyle}, {Faedi}, {Fumel}, {Gillon}, {G{\'o}mez Maqueo Chew}, {Hellier},
  {Jehin}, {Lendl}, {Maxted}, {Pepe}, {Pollacco}, {Queloz}, {S{\'e}gransan},
  {Skillen}, {Smalley}, {Smith}, {Southworth}, {Triaud}, {Turner}, {Udry}, \&
  {West}}]{2014MNRAS.445.1114A}
{Anderson}, D.~R., {Collier Cameron}, A., {Delrez}, L., {et~al.} 2014, \mnras,
  445, 1114

\bibitem[{{Astropy Collaboration} {et~al.}(2018){Astropy Collaboration},
  {Price-Whelan}, {Sip{\H o}cz}, {G{\"u}nther}, {Lim}, {Crawford}, {Conseil},
  {Shupe}, {Craig}, {Dencheva}, {Ginsburg}, {VanderPlas}, {Bradley},
  {P{\'e}rez-Su{\'a}rez}, {de Val-Borro}, {Aldcroft}, {Cruz}, {Robitaille},
  {Tollerud}, {Ardelean}, {Babej}, {Bach}, {Bachetti}, {Bakanov}, {Bamford},
  {Barentsen}, {Barmby}, {Baumbach}, {Berry}, {Biscani}, {Boquien}, {Bostroem},
  {Bouma}, {Brammer}, {Bray}, {Breytenbach}, {Buddelmeijer}, {Burke},
  {Calderone}, {Cano Rodr{\'{\i}}guez}, {Cara}, {Cardoso}, {Cheedella},
  {Copin}, {Corrales}, {Crichton}, {D'Avella}, {Deil}, {Depagne}, {Dietrich},
  {Donath}, {Droettboom}, {Earl}, {Erben}, {Fabbro}, {Ferreira}, {Finethy},
  {Fox}, {Garrison}, {Gibbons}, {Goldstein}, {Gommers}, {Greco}, {Greenfield},
  {Groener}, {Grollier}, {Hagen}, {Hirst}, {Homeier}, {Horton}, {Hosseinzadeh},
  {Hu}, {Hunkeler}, {Ivezi{\'c}}, {Jain}, {Jenness}, {Kanarek}, {Kendrew},
  {Kern}, {Kerzendorf}, {Khvalko}, {King}, {Kirkby}, {Kulkarni}, {Kumar},
  {Lee}, {Lenz}, {Littlefair}, {Ma}, {Macleod}, {Mastropietro}, {McCully},
  {Montagnac}, {Morris}, {Mueller}, {Mumford}, {Muna}, {Murphy}, {Nelson},
  {Nguyen}, {Ninan}, {N{\"o}the}, {Ogaz}, {Oh}, {Parejko}, {Parley}, {Pascual},
  {Patil}, {Patil}, {Plunkett}, {Prochaska}, {Rastogi}, {Reddy Janga},
  {Sabater}, {Sakurikar}, {Seifert}, {Sherbert}, {Sherwood-Taylor}, {Shih},
  {Sick}, {Silbiger}, {Singanamalla}, {Singer}, {Sladen}, {Sooley},
  {Sornarajah}, {Streicher}, {Teuben}, {Thomas}, {Tremblay}, {Turner},
  {Terr{\'o}n}, {van Kerkwijk}, {de la Vega}, {Watkins}, {Weaver}, {Whitmore},
  {Woillez}, {Zabalza}, \& {Astropy Contributors}}]{Astropy2018}
{Astropy Collaboration}, {Price-Whelan}, A.~M., {Sip{\H o}cz}, B.~M., {et~al.}
  2018, \aj, 156, 123

\bibitem[{{Astropy Collaboration} {et~al.}(2013){Astropy Collaboration},
  {Robitaille}, {Tollerud}, {Greenfield}, {Droettboom}, {Bray}, {Aldcroft},
  {Davis}, {Ginsburg}, {Price-Whelan}, {Kerzendorf}, {Conley}, {Crighton},
  {Barbary}, {Muna}, {Ferguson}, {Grollier}, {Parikh}, {Nair}, {Unther},
  {Deil}, {Woillez}, {Conseil}, {Kramer}, {Turner}, {Singer}, {Fox}, {Weaver},
  {Zabalza}, {Edwards}, {Azalee Bostroem}, {Burke}, {Casey}, {Crawford},
  {Dencheva}, {Ely}, {Jenness}, {Labrie}, {Lim}, {Pierfederici}, {Pontzen},
  {Ptak}, {Refsdal}, {Servillat}, \& {Streicher}}]{Astropy2013}
{Astropy Collaboration}, {Robitaille}, T.~P., {Tollerud}, E.~J., {et~al.} 2013,
  \aap, 558, A33

\bibitem[{{Benneke} \& {Seager}(2012)}]{2012ApJ...753..100B}
{Benneke}, B. \& {Seager}, S. 2012, \apj, 753, 100

\bibitem[{{Bixel} {et~al.}(2019){Bixel}, {Rackham}, {Apai}, {Espinoza},
  {L{\'o}pez-Morales}, {Osip}, {Jord{\'a}n}, {McGruder}, \&
  {Weaver}}]{2019AJ....157...68B}
{Bixel}, A., {Rackham}, B.~V., {Apai}, D., {et~al.} 2019, \aj, 157, 68

\bibitem[{{Bonomo} \& {Lanza}(2012)}]{2012A&A...547A..37B}
{Bonomo}, A.~S. \& {Lanza}, A.~F. 2012, \aap, 547, A37

\bibitem[{{Casasayas-Barris} {et~al.}(2017){Casasayas-Barris}, {Palle},
  {Nowak}, {Yan}, {Nortmann}, \& {Murgas}}]{2017A&A...608A.135C}
{Casasayas-Barris}, N., {Palle}, E., {Nowak}, G., {et~al.} 2017, \aap, 608,
  A135

\bibitem[{{Casasayas-Barris} {et~al.}(2018){Casasayas-Barris}, {Pall{\'e}},
  {Yan}, {Chen}, {Albrecht}, {Nortmann}, {Van Eylen}, {Snellen}, {Talens},
  {Gonz{\'a}lez Hern{\'a}ndez}, {Rebolo}, \& {Otten}}]{2018A&A...616A.151C}
{Casasayas-Barris}, N., {Pall{\'e}}, E., {Yan}, F., {et~al.} 2018, \aap, 616,
  A151

\bibitem[{{Casasayas-Barris} {et~al.}(2019){Casasayas-Barris}, {Pall{\'e}},
  {Yan}, {Chen}, {Kohl}, {Stangret}, {Parviainen}, {Helling}, {Watanabe},
  {Czesla}, {Fukui}, {Monta{\~n}{\'e}s-Rodr{\'\i}guez}, {Nagel}, {Narita},
  {Nortmann}, {Nowak}, {Schmitt}, \& {Zapatero Osorio}}]{2019A&A...628A...9C}
{Casasayas-Barris}, N., {Pall{\'e}}, E., {Yan}, F., {et~al.} 2019, \aap, 628,
  A9

\bibitem[{{Cepa} {et~al.}(2000){Cepa}, {Aguiar}, {Escalera},
  {Gonzalez-Serrano}, {Joven-Alvarez}, {Peraza}, {Rasilla}, {Rodriguez-Ramos},
  {Gonzalez}, {Cobos Duenas}, {Sanchez}, {Tejada}, {Bland-Hawthorn},
  {Militello}, \& {Rosa}}]{2000SPIE.4008..623C}
{Cepa}, J., {Aguiar}, M., {Escalera}, V.~G., {et~al.} 2000, in \procspie, Vol.
  4008, Optical and IR Telescope Instrumentation and Detectors, ed. M.~{Iye} \&
  A.~F. {Moorwood}, 623--631

\bibitem[{{Charbonneau} {et~al.}(2002){Charbonneau}, {Brown}, {Noyes}, \&
  {Gilliland}}]{2002ApJ...568..377C}
{Charbonneau}, D., {Brown}, T.~M., {Noyes}, R.~W., \& {Gilliland}, R.~L. 2002,
  \apj, 568, 377

\bibitem[{{Chen} {et~al.}(2020){Chen}, {Casasayas-Barris}, {Pall{\'e}}, {Yan},
  {Stangret}, {Cegla}, {Allart}, \& {Lovis}}]{2020A&A...635A.171C}
{Chen}, G., {Casasayas-Barris}, N., {Pall{\'e}}, E., {et~al.} 2020, \aap, 635,
  A171

\bibitem[{{Chen} {et~al.}(2017{\natexlab{a}}){Chen}, {Guenther}, {Pall{\'e}},
  {Nortmann}, {Nowak}, {Kunz}, {Parviainen}, \& {Murgas}}]{2017A&A...600A.138C}
{Chen}, G., {Guenther}, E.~W., {Pall{\'e}}, E., {et~al.} 2017{\natexlab{a}},
  \aap, 600, A138

\bibitem[{{Chen} {et~al.}(2017{\natexlab{b}}){Chen}, {Pall{\'e}}, {Nortmann},
  {Murgas}, {Parviainen}, \& {Nowak}}]{2017A&A...600L..11C}
{Chen}, G., {Pall{\'e}}, E., {Nortmann}, L., {et~al.} 2017{\natexlab{b}}, \aap,
  600, L11

\bibitem[{{Chen} {et~al.}(2018){Chen}, {Pall{\'e}}, {Welbanks},
  {Prieto-Arranz}, {Madhusudhan}, {Gandhi}, {Casasayas-Barris}, {Murgas},
  {Nortmann}, {Crouzet}, {Parviainen}, \& {Gandolfi}}]{2018A&A...616A.145C}
{Chen}, G., {Pall{\'e}}, E., {Welbanks}, L., {et~al.} 2018, \aap, 616, A145

\bibitem[{{Deibert} {et~al.}(2019){Deibert}, {de Mooij}, {Jayawardhana},
  {Fortney}, {Brogi}, {Rustamkulov}, \& {Tamura}}]{2019AJ....157...58D}
{Deibert}, E.~K., {de Mooij}, E.~J.~W., {Jayawardhana}, R., {et~al.} 2019, \aj,
  157, 58

\bibitem[{{Eastman} {et~al.}(2010){Eastman}, {Siverd}, \&
  {Gaudi}}]{2010PASP..122..935E}
{Eastman}, J., {Siverd}, R., \& {Gaudi}, B.~S. 2010, \pasp, 122, 935

\bibitem[{{Espinoza} \& {Jord{\'a}n}(2015)}]{2015MNRAS.450.1879E}
{Espinoza}, N. \& {Jord{\'a}n}, A. 2015, \mnras, 450, 1879

\bibitem[{{Foreman-Mackey}(2016)}]{corner}
{Foreman-Mackey}, D. 2016, The Journal of Open Source Software, 1, 24

\bibitem[{{Foreman-Mackey} {et~al.}(2013){Foreman-Mackey}, {Hogg}, {Lang}, \&
  {Goodman}}]{2013PASP..125..306F}
{Foreman-Mackey}, D., {Hogg}, D.~W., {Lang}, D., \& {Goodman}, J. 2013, \pasp,
  125, 306

\bibitem[{{Gaia Collaboration}(2018)}]{2018yCat.1345....0G}
{Gaia Collaboration}. 2018, VizieR Online Data Catalog, I/345

\bibitem[{{Gibson} {et~al.}(2012){Gibson}, {Aigrain}, {Roberts}, {Evans},
  {Osborne}, \& {Pont}}]{2012MNRAS.419.2683G}
{Gibson}, N.~P., {Aigrain}, S., {Roberts}, S., {et~al.} 2012, \mnras, 419, 2683

\bibitem[{{Hartman} {et~al.}(2009){Hartman}, {Bakos}, {Torres}, {Kov{\'a}cs},
  {Noyes}, {P{\'a}l}, {Latham}, {Sip{\H{o}}cz}, {Fischer}, {Johnson}, {Marcy},
  {Butler}, {Howard}, {Esquerdo}, {Sasselov}, {Kov{\'a}cs}, {Stefanik},
  {Fernandez}, {L{\'a}z{\'a}r}, {Papp}, \& {S{\'a}ri}}]{2009ApJ...706..785H}
{Hartman}, J.~D., {Bakos}, G.~{\'A}., {Torres}, G., {et~al.} 2009, \apj, 706,
  785

\bibitem[{{Heng} \& {Kitzmann}(2017)}]{2017MNRAS.470.2972H}
{Heng}, K. \& {Kitzmann}, D. 2017, \mnras, 470, 2972

\bibitem[{{H{\o}g} {et~al.}(2000){H{\o}g}, {Fabricius}, {Makarov}, {Urban},
  {Corbin}, {Wycoff}, {Bastian}, {Schwekendiek}, \&
  {Wicenec}}]{2000A&A...355L..27H}
{H{\o}g}, E., {Fabricius}, C., {Makarov}, V.~V., {et~al.} 2000, \aap, 355, L27

\bibitem[{{Horne}(1986)}]{1986PASP...98..609H}
{Horne}, K. 1986, \pasp, 98, 609

\bibitem[{Hunter(2007)}]{Hunter2007}
Hunter, J.~D. 2007, Computing In Science \& Engineering, 9, 90

\bibitem[{{Husser} {et~al.}(2013){Husser}, {Wende-von Berg}, {Dreizler},
  {Homeier}, {Reiners}, {Barman}, \& {Hauschildt}}]{2013A&A...553A...6H}
{Husser}, T.~O., {Wende-von Berg}, S., {Dreizler}, S., {et~al.} 2013, \aap,
  553, A6

\bibitem[{{Ioannidis} \& {Schmitt}(2016)}]{2016A&A...594A..41I}
{Ioannidis}, P. \& {Schmitt}, J.~H.~M.~M. 2016, \aap, 594, A41

\bibitem[{Jones {et~al.}(2001)Jones, Oliphant, Peterson, {et~al.}}]{Jones2001}
Jones, E., Oliphant, T., Peterson, P., {et~al.} 2001, {SciPy}: Open source
  scientific tools for {Python}, [Online; accessed <today>]

\bibitem[{{Jord{\'a}n} {et~al.}(2013){Jord{\'a}n}, {Espinoza}, {Rabus},
  {Eyheramendy}, {Sing}, {D{\'e}sert}, {Bakos}, {Fortney}, {L{\'o}pez-Morales},
  {Maxted}, {Triaud}, \& {Szentgyorgyi}}]{2013ApJ...778..184J}
{Jord{\'a}n}, A., {Espinoza}, N., {Rabus}, M., {et~al.} 2013, \apj, 778, 184

\bibitem[{{Kempton} {et~al.}(2017){Kempton}, {Lupu}, {Owusu-Asare}, {Slough},
  \& {Cale}}]{2017PASP..129d4402K}
{Kempton}, E. M.~R., {Lupu}, R., {Owusu-Asare}, A., {Slough}, P., \& {Cale}, B.
  2017, \pasp, 129, 044402

\bibitem[{{Kreidberg}(2015)}]{2015PASP..127.1161K}
{Kreidberg}, L. 2015, \pasp, 127, 1161

\bibitem[{{Kreidberg} {et~al.}(2014){Kreidberg}, {Bean}, {D{\'e}sert},
  {Benneke}, {Deming}, {Stevenson}, {Seager}, {Berta-Thompson}, {Seifahrt}, \&
  {Homeier}}]{2014Natur.505...69K}
{Kreidberg}, L., {Bean}, J.~L., {D{\'e}sert}, J.-M., {et~al.} 2014, \nat, 505,
  69

\bibitem[{{Kurucz}(1979)}]{1979ApJS...40....1K}
{Kurucz}, R.~L. 1979, \apjs, 40, 1

\bibitem[{{Lecavelier Des Etangs} {et~al.}(2008){Lecavelier Des Etangs},
  {Pont}, {Vidal-Madjar}, \& {Sing}}]{2008A&A...481L..83L}
{Lecavelier Des Etangs}, A., {Pont}, F., {Vidal-Madjar}, A., \& {Sing}, D.
  2008, \aap, 481, L83

\bibitem[{{Mancini} {et~al.}(2018){Mancini}, {Esposito}, {Covino},
  {Southworth}, {Biazzo}, {Bruni}, {Ciceri}, {Evans}, {Lanza}, {Poretti},
  {Sarkis}, {Smith}, {Brogi}, {Affer}, {Benatti}, {Bignamini}, {Boccato},
  {Bonomo}, {Borsa}, {Carleo}, {Claudi}, {Cosentino}, {Damasso}, {Desidera},
  {Giacobbe}, {Gonz{\'a}lez-{\'A}lvarez}, {Gratton}, {Harutyunyan}, {Leto},
  {Maggio}, {Malavolta}, {Maldonado}, {Martinez-Fiorenzano}, {Masiero},
  {Micela}, {Molinari}, {Nascimbeni}, {Pagano}, {Pedani}, {Piotto}, {Rainer},
  {Scand ariato}, {Smareglia}, {Sozzetti}, {Andreuzzi}, \&
  {Henning}}]{2018A&A...613A..41M}
{Mancini}, L., {Esposito}, M., {Covino}, E., {et~al.} 2018, \aap, 613, A41

\bibitem[{{Mandel} \& {Agol}(2002)}]{2002ApJ...580L.171M}
{Mandel}, K. \& {Agol}, E. 2002, \apjl, 580, L171

\bibitem[{{May} {et~al.}(2020){May}, {Gardner}, {Rauscher}, \&
  {Monnier}}]{2020AJ....159....7M}
{May}, E.~M., {Gardner}, T., {Rauscher}, E., \& {Monnier}, J.~D. 2020, \aj,
  159, 7

\bibitem[{{May} {et~al.}(2018){May}, {Zhao}, {Haidar}, {Rauscher}, \&
  {Monnier}}]{2018AJ....156..122M}
{May}, E.~M., {Zhao}, M., {Haidar}, M., {Rauscher}, E., \& {Monnier}, J.~D.
  2018, \aj, 156, 122

\bibitem[{{McCullough} {et~al.}(2014){McCullough}, {Crouzet}, {Deming}, \&
  {Madhusudhan}}]{2014ApJ...791...55M}
{McCullough}, P.~R., {Crouzet}, N., {Deming}, D., \& {Madhusudhan}, N. 2014,
  \apj, 791, 55

\bibitem[{{Murgas} {et~al.}(2019){Murgas}, {Chen}, {Pall{\'e}}, {Nortmann}, \&
  {Nowak}}]{2019A&A...622A.172M}
{Murgas}, F., {Chen}, G., {Pall{\'e}}, E., {Nortmann}, L., \& {Nowak}, G. 2019,
  \aap, 622, A172

\bibitem[{{Nortmann} {et~al.}(2018){Nortmann}, {Pall{\'e}}, {Salz},
  {Sanz-Forcada}, {Nagel}, {Alonso-Floriano}, {Czesla}, {Yan}, {Chen},
  {Snellen}, {Zechmeister}, {Schmitt}, {L{\'o}pez-Puertas}, {Casasayas-Barris},
  {Bauer}, {Amado}, {Caballero}, {Dreizler}, {Henning}, {Lamp{\'o}n}, {Montes},
  {Molaverdikhani}, {Quirrenbach}, {Reiners}, {Ribas}, {S{\'a}nchez-L{\'o}pez},
  {Schneider}, \& {Zapatero Osorio}}]{2018Sci...362.1388N}
{Nortmann}, L., {Pall{\'e}}, E., {Salz}, M., {et~al.} 2018, Science, 362, 1388

\bibitem[{{Oshagh} {et~al.}(2014){Oshagh}, {Santos}, {Ehrenreich},
  {Haghighipour}, {Figueira}, {Santerne}, \& {Montalto}}]{2014A&A...568A..99O}
{Oshagh}, M., {Santos}, N.~C., {Ehrenreich}, D., {et~al.} 2014, \aap, 568, A99

\bibitem[{P\'erez \& Granger(2007)}]{PER-GRA:2007}
P\'erez, F. \& Granger, B.~E. 2007, Computing in Science and Engineering, 9, 21

\bibitem[{{Pont} {et~al.}(2013){Pont}, {Sing}, {Gibson}, {Aigrain}, {Henry}, \&
  {Husnoo}}]{2013MNRAS.432.2917P}
{Pont}, F., {Sing}, D.~K., {Gibson}, N.~P., {et~al.} 2013, \mnras, 432, 2917

\bibitem[{{Rackham} {et~al.}(2017){Rackham}, {Espinoza}, {Apai},
  {L{\'o}pez-Morales}, {Jord{\'a}n}, {Osip}, {Lewis}, {Rodler}, {Fraine},
  {Morley}, \& {Fortney}}]{2017ApJ...834..151R}
{Rackham}, B., {Espinoza}, N., {Apai}, D., {et~al.} 2017, \apj, 834, 151

\bibitem[{{Rackham} {et~al.}(2018){Rackham}, {Apai}, \&
  {Giampapa}}]{2018ApJ...853..122R}
{Rackham}, B.~V., {Apai}, D., \& {Giampapa}, M.~S. 2018, \apj, 853, 122

\bibitem[{{Rackham} {et~al.}(2019){Rackham}, {Apai}, \&
  {Giampapa}}]{2019AJ....157...96R}
{Rackham}, B.~V., {Apai}, D., \& {Giampapa}, M.~S. 2019, \aj, 157, 96

\bibitem[{{Rasmussen} \& {Williams}(2010)}]{2006GP}
{Rasmussen}, C. \& {Williams}, C. 2010, the MIT Press, 122, 935

\bibitem[{{Redfield} {et~al.}(2008){Redfield}, {Endl}, {Cochran}, \&
  {Koesterke}}]{2008ApJ...673L..87R}
{Redfield}, S., {Endl}, M., {Cochran}, W.~D., \& {Koesterke}, L. 2008, \apjl,
  673, L87

\bibitem[{{S{\'a}nchez} {et~al.}(2012){S{\'a}nchez}, {Aguiar-Gonz{\'a}lez},
  {Barreto}, {Becerril}, {Bland-Hawthorn}, {Bongiovanni}, {Cepa}, {Correa},
  {Chapa}, {Ederoclite}, {Espejo}, {Farah}, {Fragoso}, {Fern{\'a}ndez},
  {Flores}, {Fuentes}, {Gago}, {Garfias}, {Gigante}, {Gonz{\'a}lez},
  {Gonz{\'a}lez-Escalera}, {Hern{\'a}ndez}, {Hernandez}, {Herrera}, {Herrera},
  {Joven}, {Langarica}, {Lara}, {L{\'o}pez}, {L{\'o}pez}, {Militellon},
  {Moreno}, {Peraza}, {P{\'e}rez}, {P{\'e}rez}, {Rasilla}, {Rosich}, {Tejada},
  {Tinoco}, {Vaz}, \& {Villegas}}]{2012SPIE.8446E..4TS}
{S{\'a}nchez}, B., {Aguiar-Gonz{\'a}lez}, M., {Barreto}, R., {et~al.} 2012, in
  \procspie, Vol. 8446, Ground-based and Airborne Instrumentation for Astronomy
  IV, 84464T

\bibitem[{{Seidel} {et~al.}(2019){Seidel}, {Ehrenreich}, {Wyttenbach},
  {Allart}, {Lendl}, {Pino}, {Bourrier}, {Cegla}, {Lovis}, {Barrado},
  {Bayliss}, {Astudillo-Defru}, {Deline}, {Fisher}, {Heng}, {Joseph}, {Lavie},
  {Melo}, {Pepe}, {S{\'e}gransan}, \& {Udry}}]{2019A&A...623A.166S}
{Seidel}, J.~V., {Ehrenreich}, D., {Wyttenbach}, A., {et~al.} 2019, \aap, 623,
  A166

\bibitem[{{Sing} {et~al.}(2016){Sing}, {Fortney}, {Nikolov}, {Wakeford},
  {Kataria}, {Evans}, {Aigrain}, {Ballester}, {Burrows}, {Deming},
  {D{\'e}sert}, {Gibson}, {Henry}, {Huitson}, {Knutson}, {Lecavelier Des
  Etangs}, {Pont}, {Showman}, {Vidal-Madjar}, {Williamson}, \&
  {Wilson}}]{2016Natur.529...59S}
{Sing}, D.~K., {Fortney}, J.~J., {Nikolov}, N., {et~al.} 2016, \nat, 529, 59

\bibitem[{{Sing} {et~al.}(2012){Sing}, {Huitson}, {Lopez-Morales}, {Pont},
  {D{\'e}sert}, {Ehrenreich}, {Wilson}, {Ballester}, {Fortney}, {Lecavelier des
  Etangs}, \& {Vidal-Madjar}}]{2012MNRAS.426.1663S}
{Sing}, D.~K., {Huitson}, C.~M., {Lopez-Morales}, M., {et~al.} 2012, \mnras,
  426, 1663

\bibitem[{{Sing} {et~al.}(2011){Sing}, {Pont}, {Aigrain}, {Charbonneau},
  {D{\'e}sert}, {Gibson}, {Gilliland}, {Hayek}, {Henry}, {Knutson}, {Lecavelier
  Des Etangs}, {Mazeh}, \& {Shporer}}]{2011MNRAS.416.1443S}
{Sing}, D.~K., {Pont}, F., {Aigrain}, S., {et~al.} 2011, \mnras, 416, 1443

\bibitem[{{Speagle}(2020)}]{2020MNRAS.493.3132S}
{Speagle}, J.~S. 2020, \mnras, 493, 3132

\bibitem[{{Todorov} {et~al.}(2019){Todorov}, {D{\'e}sert}, {Huitson}, {Bean},
  {Panwar}, {de Matos}, {Stevenson}, {Fortney}, \&
  {Bergmann}}]{2019A&A...631A.169T}
{Todorov}, K.~O., {D{\'e}sert}, J.-M., {Huitson}, C.~M., {et~al.} 2019, \aap,
  631, A169

\bibitem[{{Tsiaras} {et~al.}(2018){Tsiaras}, {Waldmann}, {Zingales},
  {Rocchetto}, {Morello}, {Damiano}, {Karpouzas}, {Tinetti}, {McKemmish},
  {Tennyson}, \& {Yurchenko}}]{2018AJ....155..156T}
{Tsiaras}, A., {Waldmann}, I.~P., {Zingales}, T., {et~al.} 2018, \aj, 155, 156

\bibitem[{{van der Walt} {et~al.}(2011){van der Walt}, {Colbert}, \&
  {Varoquaux}}]{vanderwalt2011}
{van der Walt}, S., {Colbert}, S.~C., \& {Varoquaux}, G. 2011, Computing in
  Science and Engineering, 13, 22

\bibitem[{{Weaver} {et~al.}(2020){Weaver}, {L{\'o}pez-Morales}, {Espinoza},
  {Rackham}, {Osip}, {Apai}, {Jord{\'a}n}, {Bixel}, {Lewis}, {Alam}, {Kirk},
  {McGruder}, {Rodler}, \& {Fienco}}]{2020AJ....159...13W}
{Weaver}, I.~C., {L{\'o}pez-Morales}, M., {Espinoza}, N., {et~al.} 2020, \aj,
  159, 13

\bibitem[{{Wong} {et~al.}(2020){Wong}, {Benneke}, {Gao}, {Knutson}, {Chachan},
  {Henry}, {Deming}, {Kataria}, {Lee}, {Nikolov}, {Sing}, {Ballester},
  {Baskin}, {Wakeford}, \& {Williamson}}]{2020AJ....159..234W}
{Wong}, I., {Benneke}, B., {Gao}, P., {et~al.} 2020, \aj, 159, 234

\bibitem[{{Wood} {et~al.}(2011){Wood}, {Maxted}, {Smalley}, \&
  {Iro}}]{2011MNRAS.412.2376W}
{Wood}, P.~L., {Maxted}, P.~F.~L., {Smalley}, B., \& {Iro}, N. 2011, \mnras,
  412, 2376

\bibitem[{{Wyttenbach} {et~al.}(2017){Wyttenbach}, {Lovis}, {Ehrenreich},
  {Bourrier}, {Pino}, {Allart}, {Astudillo-Defru}, {Cegla}, {Heng}, {Lavie},
  {Melo}, {Murgas}, {Santerne}, {S{\'e}gransan}, {Udry}, \&
  {Pepe}}]{2017A&A...602A..36W}
{Wyttenbach}, A., {Lovis}, C., {Ehrenreich}, D., {et~al.} 2017, \aap, 602, A36

\bibitem[{{Zhang} {et~al.}(2019){Zhang}, {Chachan}, {Kempton}, \&
  {Knutson}}]{2019PASP..131c4501Z}
{Zhang}, M., {Chachan}, Y., {Kempton}, E. M.~R., \& {Knutson}, H.~A. 2019,
  \pasp, 131, 034501

\end{thebibliography}
%



\begin{appendix}
  \label{sec:appendix}

\section{Additional tables and figures}
Here we present the transmission spectrum measurement tables, the correlation plots of the MCMC fitting procedure of spots and faculae modeling, and the transmission spectrum around the Na doublet region. 
 
\begin{table}
 \caption[]{Results for the planet-to-star radius ratio for each of the 19 spectroscopic channels.}
\begin{center}
\def\tol#1#2#3{\hbox{\rule{0pt}{15pt}${#1}^{{#2}}_{{#3}}$}}
\setlength{\tabcolsep}{1.5mm}
\begin{tabular}{c c c}
\hline\hline
Center (nm) &  Width (nm) &   $R_{\rm{p}}/R_\star$  \\\hline  
        525      &  20 &       \tol{0.132403}{+0.000446}{-0.000539} \\
        545      &  20 &         \tol{0.133965}{+0.000386}{-0.000374} \\
        565      &  20 &         \tol{0.133097}{+0.000351}{-0.000455} \\
        585      &  20 &         \tol{0.132887}{+0.000260}{-0.000249} \\
        605      &  20 &         \tol{0.132740}{+0.000352}{-0.000355} \\
        625      &  20 &         \tol{0.132589}{+0.001475}{-0.001080} \\
        645      &  20 &         \tol{0.133365}{+0.000603}{-0.000541} \\
        665      &  20 &         \tol{0.132099}{+0.000234}{-0.000226} \\
        684      &  20 &         \tol{0.132304}{+0.000255}{-0.000231} \\
        705      &  20 &         \tol{0.132087}{+0.000194}{-0.000188} \\
        725      &  20 &         \tol{0.132943}{+0.000728}{-0.000855} \\
        745      &  20 &         \tol{0.132043}{+0.000310}{-0.000307} \\
        775      &  20 &         \tol{0.131464}{+0.000239}{-0.000221} \\
        795      &  20 &         \tol{0.131294}{+0.000427}{-0.000226} \\
        815      &  20 &         \tol{0.132059}{+0.000242}{-0.000237} \\
        835      &  20 &         \tol{0.131341}{+0.000283}{-0.000310} \\
        855      &  20 &         \tol{0.131626}{+0.000242}{-0.000238} \\
        875      &  20 &         \tol{0.131268}{+0.000255}{-0.000252} \\
        895      &  20 &         \tol{0.131431}{+0.000346}{-0.000309} \\
        \hline
\end{tabular}
\label{tab:specrprs}  
\end{center}
\end{table}
  
\begin{table}
 \caption[]{Results for the planet-to-star radius ratio for each of the 11 spectroscopic channels around the Na {\sc i} doublet.}
\begin{center}
\def\tol#1#2#3{\hbox{\rule{0pt}{15pt}${#1}^{{#2}}_{{#3}}$}}
\setlength{\tabcolsep}{1.5mm}
\begin{tabular}{c c c}
\hline\hline
Center (nm) &  Width (nm) &   $R_{\rm{p}}/R_\star$  \\\hline  
564.3 & 5 & \tol{0.133100}{+0.000501}{-0.000473} \\
569.3 & 5 & \tol{0.133121}{+0.000431}{-0.000484} \\
574.3 & 5 & \tol{0.132683}{+0.000483}{-0.000527} \\
579.3 & 5 & \tol{0.133343}{+0.000419}{-0.000425} \\
584.3 & 5 & \tol{0.133790}{+0.000467}{-0.000444} \\
589.3 & 5 & \tol{0.129845}{+0.001861}{-0.002393} \\
594.3 & 5 & \tol{0.132244}{+0.000484}{-0.000457} \\
599.3 & 5 & \tol{0.131914}{+0.000418}{-0.000416} \\
604.3 & 5 & \tol{0.132476}{+0.000442}{-0.000447} \\
609.3 & 5 & \tol{0.133244}{+0.001299}{-0.001218} \\
614.3 & 5 & \tol{0.130494}{+0.002538}{-0.001250} \\
        \hline
\end{tabular}
\label{tab:specrprsNa}  
\end{center}
\end{table}

   \begin{figure*}
   \centering
   \includegraphics[width=\textwidth]{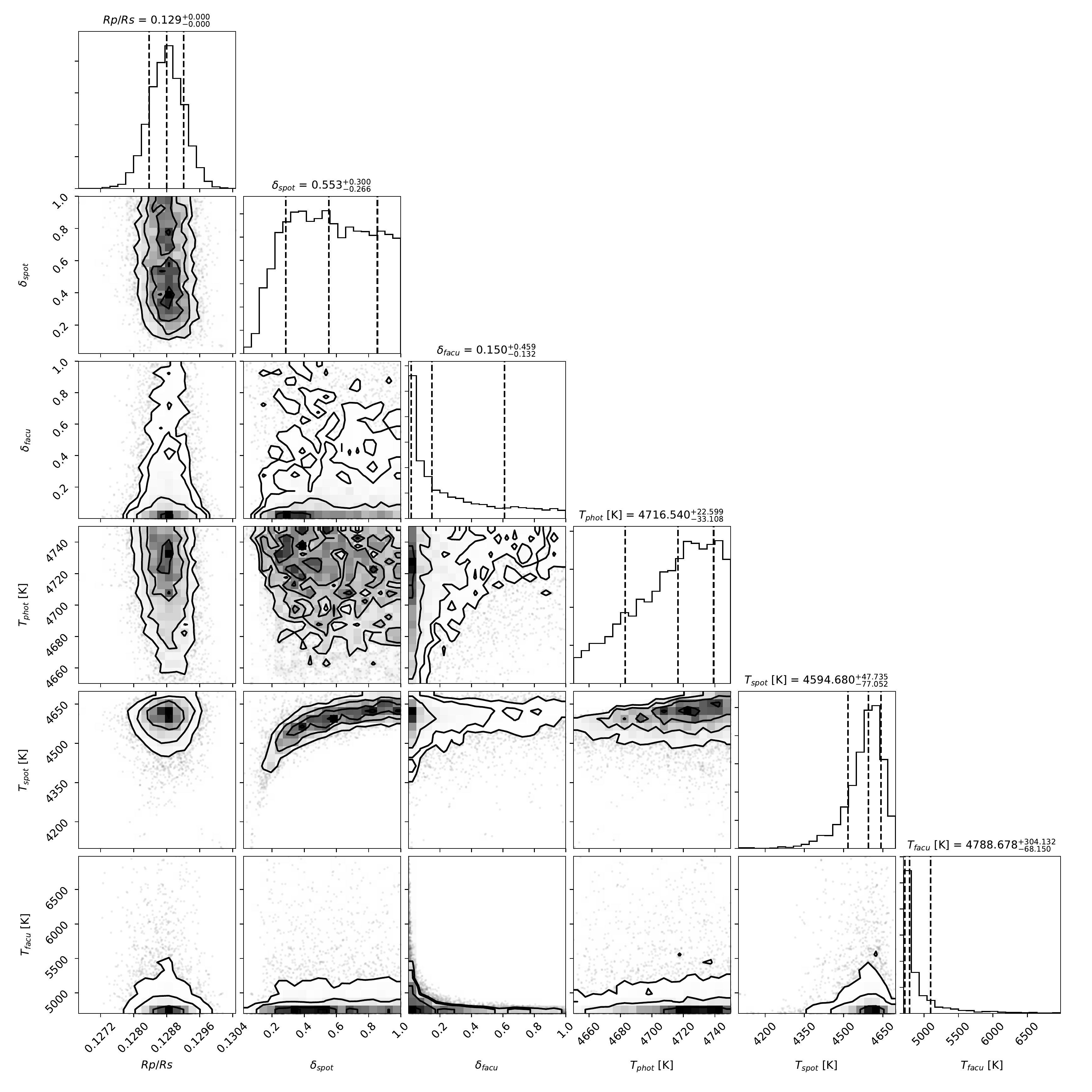}
   \caption{Correlation plot of fitted parameters for the spot and faculae coverage model.}
   \label{FigSpotCorner}%
   \end{figure*}
   
  \begin{figure*}
    \centering
    \includegraphics[width=\hsize]{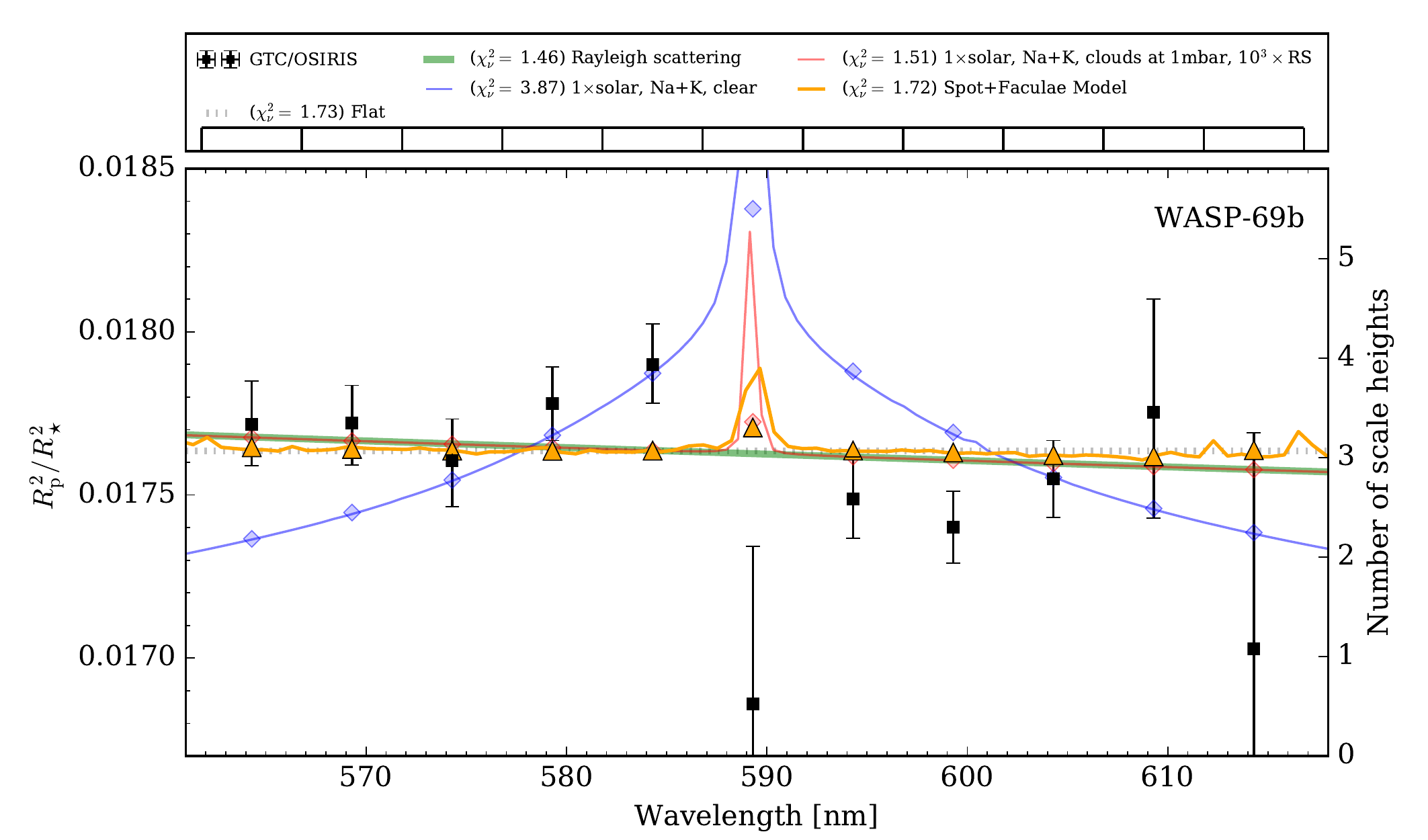}
    \caption{Transmission spectrum of WASP-69b around the Na {\sc i} doublet (black squares) compared to several atmosphere models. In blue, a model for a clear atmosphere and solar abundance is shown. The gray dashed line indicates the model of flat transmission spectrum. The slope expected in an atmosphere affected solely by Rayleigh scattering is shown in green. In red,  we show the model of an atmosphere with clouds at 1 mbar and Rayleigh scattering enhanced by a factor of 1000. In orange, the best fitted unocculted spots and faculae model is shown.}
    \label{fig:Naline}
  \end{figure*}

\end{appendix}

\end{document}